\documentclass[preprint,showpacs,singlecolumn,superscriptaddress]{revtex4-1}
\usepackage{graphicx}
\usepackage{tabularx}
\usepackage{color}
\usepackage{amsmath}
\usepackage{comment}
\newcommand{\be}{\begin{equation}}
\newcommand{\ee}{\end{equation}}
\newcommand{\bea}{\begin{eqnarray}}
\newcommand{\eea}{\end{eqnarray}}
\usepackage{dcolumn}
\usepackage{bm}
\usepackage{epsf}
\usepackage{subfigure}
\usepackage{epstopdf}%
\usepackage{amsfonts}

\begin{document}

\title{Tunable photonic cavity coupled to a voltage-biased double quantum dot system: Diagrammatic NEGF approach}

\author{Bijay Kumar Agarwalla}
\affiliation{Chemical Physics Theory Group, Department of Chemistry,
and Centre for Quantum Information and Quantum Control,
University of Toronto, 80 Saint George St., Toronto, Ontario, Canada M5S 3H6}

\author{Manas Kulkarni}
\affiliation{Department of Physics, New York City College of Technology, The City University of New York, Brooklyn, New York 11201, USA}

\author{Shaul Mukamel}
\affiliation{Department of Chemistry, University of California, Irvine, California 92697, USA}

\author{Dvira Segal}
\affiliation{Chemical Physics Theory Group, Department of Chemistry,
and Centre for Quantum Information and Quantum Control,
University of Toronto, 80 Saint George St., Toronto, Ontario, Canada M5S 3H6}

\date{\today}

\begin{abstract}

We investigate gain in microwave photonic cavities coupled to voltage-biased  double quantum dot systems with an 
arbitrary strong dot-lead coupling and with a \textit{Holstein-like} light-matter interaction,
by adapting the diagrammatic Keldysh nonequilibrium Green's function approach.
We compute out-of-equilibrium properties of the cavity : its
transmission, phase response, mean photon number, power spectrum, and spectral function.
We show that by the careful engineering of these hybrid light-matter systems, one can achieve a significant amplification of the optical signal
with the voltage-biased electronic system serving as a gain medium.
We also study
the steady state current across the device, identifying elastic and inelastic tunnelling processes which involve
the cavity mode.
Our results show how recent advances in quantum electronics 
can be exploited to build hybrid light-matter systems that behave as single-atom amplifiers and photon source devices.  
The diagrammatic Keldysh approach is primarily discussed for a cavity-coupled double quantum dot architecture, 
but it is generalizable to other hybrid light-matter systems.





\end{abstract}

\maketitle

\section{Introduction}
\label{intro}

Recent years have seen a significant progress in probing and controlling hybrid light-matter systems  
at the interface of quantum optics and condensed matter physics \cite{RevNori,KLH2015, PNAS1,Hincks_control}. 
Few examples of hybrid quantum systems include
cavity-Quantum Electrodynamics (c-QED) arrays \cite{nat_phys_rev,KLH2015,underwood_PhysRevA.86.023837,sskj}, 
cold atoms coupled to light \cite{esslinger10, kulkarniprl,Krinner2015}, 
optomechanical devices \cite{cavop,cavop1} and cavity-coupled quantum dots 
\cite{Petersson2012, kontoskondo,kontosnatcom, petta2014, Kulkarni2014,KLH2015, enslinprl}.  
The motivation for this paper is a class of recent experiments where quantum dots have been integrated with superconducting resonators, 
accomplishing sufficiently strong charge-cavity coupling of $g$ $\sim 50-200$ MHz 
\cite{Frey2012, Petersson2012, Toida2013, Viennot2013, Deng2013, gpg}. 
Such quantum-dot circuit QED systems (QD-cQED) 
offer a rich platform for studying non-equilibrium open quantum systems at the interface of 
quantum optics and mesoscopic solid-state physics. 
Experiments are versatile, with a highly tunable window of parameters. 
Recent breakthroughs in such devices include the observation of
photon emission proceeding via the DC transport of electrons \cite{petta2014}, and the
realization of microwave lasers (\textit{masers}) \cite{pettamaser,pettamaserscience}. 

Despite ongoing theoretical advances in describing QD-cQED systems \cite{KLH2015,Kulkarni2014,marco14,simon16,hartlekulkarniprb},  
there is a compelling need for adapting well-established techniques of non-equilibrium and condensed matter physics, 
e.g, the diagrammatic non-equilibrium Green's function (NEGF) method, 
to explore the rich physics of these highly tunable and versatile hybrid light-matter systems. 
Existing literature typically treats electron-lead coupling in a perturbative manner (such as the Born-Markov approximation) 
\cite{Kulkarni2014,xu2013quantum,xu2013full,jin2011lasing,andre2009few,schon16}, further
enforcing the source-drain voltage to be very high, thereby incorporating only sequential, uni-directional electron transfer across the dots.
Using such approaches one potentially misses important features in the optical and electronic signals, the result of {\it finite} bias voltage
and {\it strong} dot-lead couplings.
Moreover, approximate methods such as the Markovian-secular quantum master equation or mean field calculations are often uncontrolled and non-transparent. 
It is therefore of a great importance to introduce a systematic approach that allows for an arbitrary dot-lead coupling, 
(especially since experiments allow tunability from weak to strong dot-lead coupling), 
handles finite source-drain bias voltage,
and treats light-matter coupling in a systematic (even if perturbative) manner.  
The diagrammatic NEGF approach \cite{negf1,negf2,negf3,harbolamukamel} is perfectly suited for this purpose. It
allows us to simulate present cutting-edge experimental realizations of quantum-dot circuit-QED systems,
and furthermore foresee new effects.

\begin{figure} [pb]
\includegraphics[width=8.5cm]{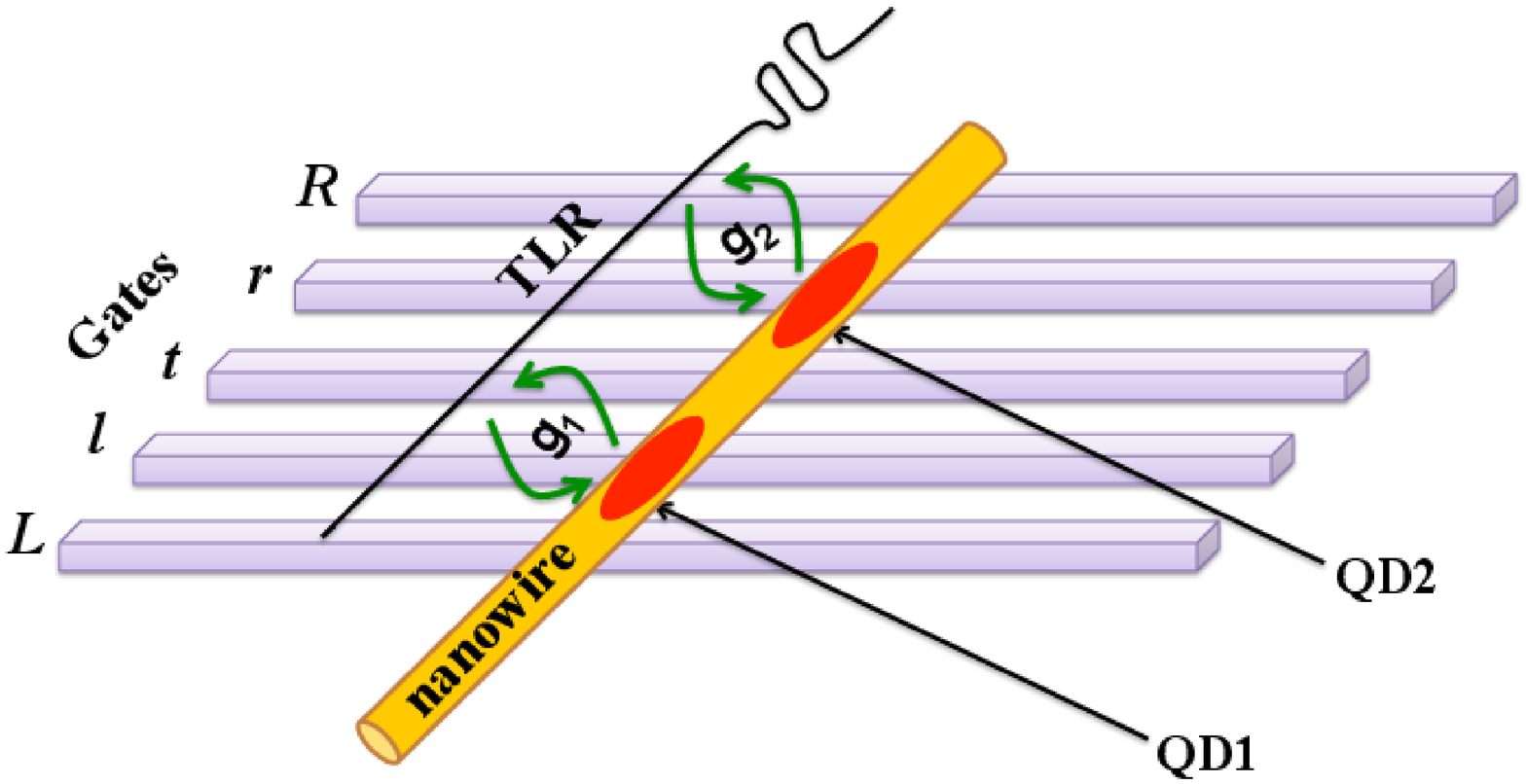}\qquad
\includegraphics[width=7cm]{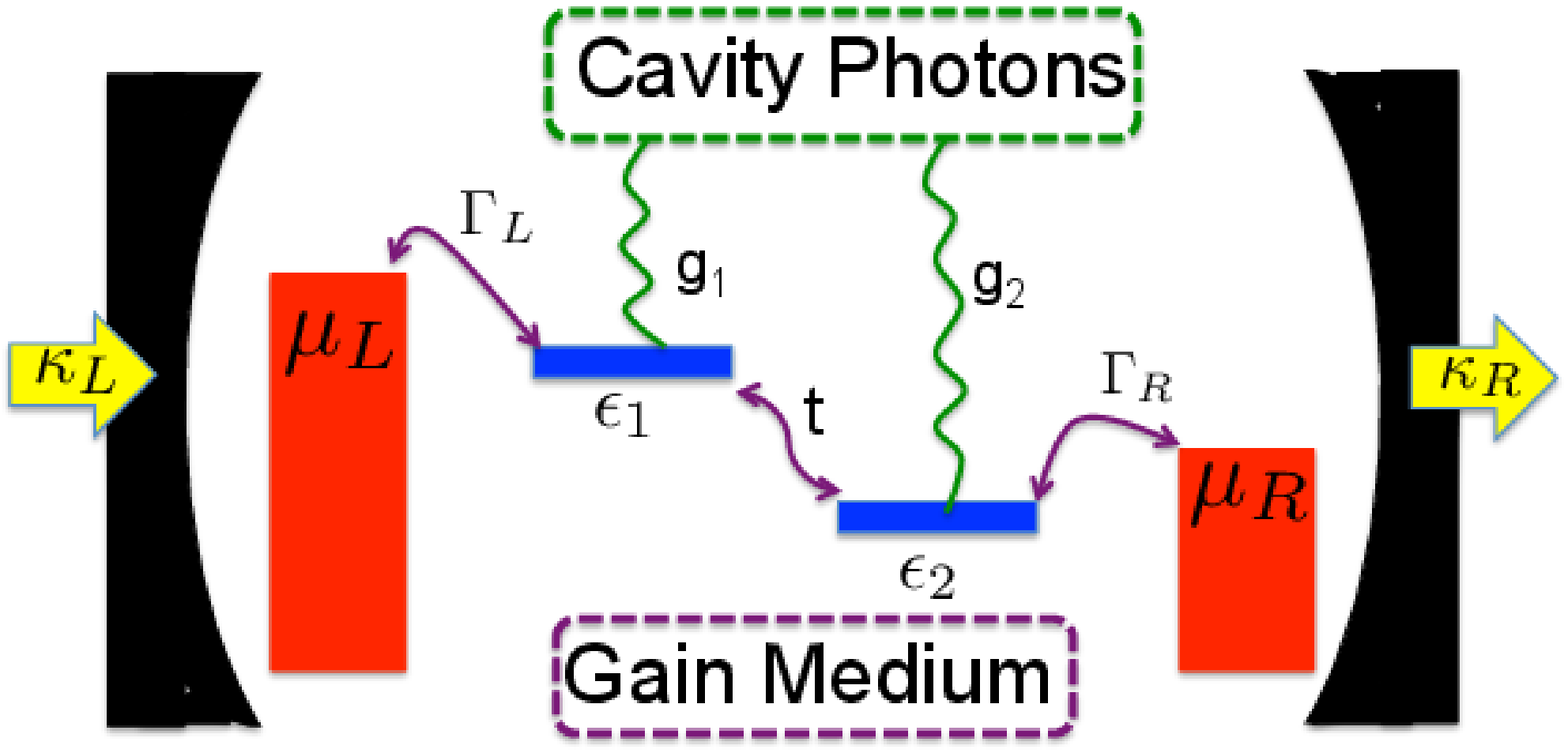}
\caption{(Left) Schematics of an implementation of a biased double-quantum-dot circuit-QED
setup: The electronic-mesoscopic system is integrated with a superconducting transmission line resonator (TLR). 
The coupling between the microwave photons of the TLR and the electronic levels of each quantum
dot (QD), embedded in the nanowire, is described by a Holstein interaction of strength ($g_1,g_2$). 
The DQD is driven out of equilibrium by
the application of a finite source-drain bias $\Delta \mu =\mu_L -\mu_R$.
Tunnelling rates between the dots and the electron leads ($\Gamma_L,\Gamma_R$) and in between the dots ($t$) can be tuned via gate-controlled tunnel-barriers.  
Typical experimentally relevant values are given in the Table. 
(Right) A schematic representation of the model. The mesoscopic system is effectively housed in the microwave cavity. Cavity photons are coupled to the input and output ports with decay rates $\kappa_{L(R)}$. The arrows inside the cavity represent tunnelling processes. Wavy lines indicate the light-matter coupling.}
\label{schematic}
\end{figure}

Our model includes two electronic levels corresponding to two quantum dots (DQD), each coupled to a primary microwave photon mode (\text{cavity photons}). 
This primary mode is coupled to left and right transmission lines, mimicking the \textit{openness} of the cavity. 
A source-drain bias is applied across the DQD system, inducing DC electric current. 
This system can serve as a testbed for understanding the intricate interplay between light (cavity) and matter (voltage-biased DQD) degrees of freedom,
specifically, in a non-equilibrium situation.
For a schematic representation, see Fig. \ref{schematic}.


The paper in organized as follows. 
In Sec.~\ref{model} we introduce our model.
In Sec.~\ref{meanPh} we study the optical properties of the cavity, 
namely, the mean photon number, power spectrum and the spectral function, transmission coefficient, and phase response. We use the Keldysh NEGF method while relying on the random-phase-approximation (RPA), 
which is crucial for respecting symmetry conditions. 
Numerical simulations demonstrate the potential to use the system and develop novel quantum devices such as microwave amplifiers, 
photon sources and diodes/rectifiers. In Sec.~\ref{CGF-NEGF} we focus on the 
electronic part of the model, and demonstrate the influence of the cavity mode on the electric charge current. 
We summarize our work and provide an outlook in Sec. \ref{Summary}.


\section{Model}
\label{model}

The double quantum dot setup is placed between two metal leads composed of non-interacting electrons. 
Electron transfer between the dots takes place via direct tunnelling. 
Each dot is further coupled to a microwave cavity mode, designated as the ``primary photon". 
This mode is coupled to two transmission lines, namely input and output ports. 
The total Hamiltonian is (we set $\hbar\!=\!k_B\!=\!e\!=\!1$ throughout the paper),
\bea
\hat H_T= \hat H_{el} + \hat H_{ph} + \hat H_{el-ph},
\label{eq:Htotal}
\eea
with
\bea
\hat H_{el}&=& {\epsilon}_1 \hat c_1^{\dagger} \hat c_1 + {\epsilon}_2 \hat c_2^{\dagger} \hat c_2 + t (\hat c_1^{\dagger} \hat c_2 + \hat c_2^{\dagger} \hat c_1)
+ \sum_{l\in L} {\epsilon}_l \hat c_l^{\dagger} \hat c_l \nonumber\\
&+& \sum_{r\in R} {\epsilon}_r \hat c_r^{\dagger} \hat c_r
+ \sum_{l \in L } v_{l} (\hat c_l^{\dagger} \hat c_1 \!+\! \hat c_1^{\dagger} \hat c_l) \!+\! \sum_{r \in R} v_{r} (\hat c_r^{\dagger} \hat c_2 \!+\! \hat c_{2}^{\dagger} \hat c_r),\,\,\,
\label{eq:Hel}
\eea
where $\epsilon_{1}, \epsilon_{2}$ are the site energies of the DQD, coupled to
the left $L$ and right $R$ metal leads by real-valued hopping elements $v_l$ and $v_r$, respectively.
$\hat c^{\dagger}$ and $\hat c$ are fermionic creation and annihilation operators for the respective dots.
$\hat H_{ph}$ is the Hamiltonian for the photonic degrees of freedom. It
consists of the primary photon of frequency $\omega_0$ and the secondary photon baths as two long transmission lines ($K=L,R$) with a symmetric coupling $\nu_j$
\be
\hat H_{ph} = \omega_0  \hat a^{\dagger} \hat a +
\sum_{j\in K} \omega_{jK} \hat a_{jK}^{\dagger} \hat a_{jK} + \sum_{j\in K} \nu_{j} \,\hat a_{jK}^{\dagger} \, \hat {a} + h.c.
\label{eq:Hvib}
\ee
Here, $\hat a(\hat a^{\dagger})$ and $\hat a_{jK}(\hat a_{jK}^{\dagger})$ are bosonic annihilation (creation) operators for
the cavity mode and the two transmission lines. 

The interaction between electrons in the junction and the primary optical mode is given by 
\be
\hat H_{el-ph} = [g_1 \hat{n}_1  + g_2  \hat{n}_2 ]  (\hat a^{\dagger} + \hat a),
\label{eq:Hevib}
\ee
with $\hat{n}_{\alpha}=\hat c_{\alpha}^{\dagger} \hat c_{\alpha}$ as the level number operator 
and $g_{\alpha}$ the coupling strength, $\alpha=1,2$.
For $g_1\!=\!-g_2$, $\hat{H}_{{el-ph}}$ describes the interaction between the microwave photon and the dipole moment of excess electrons in the DQD.
Experimentally relevant parameters  are given in the Table. 
$\kappa\!=\!\kappa_L\!=\!\kappa_R$ is the decay rate of the cavity mode per port. In the wide-band limit we define it as $\kappa= 2 \pi \rho |\nu|^2$, where $\rho$ is the bath density of states and $\nu$ being the average coupling between the cavity and the bath modes.


\label{tab:title}
\begin{center}
Table:  Typical Parameter values from experiments (Refs.~\onlinecite{petta2014,Kulkarni2014})
\begin{tabularx}{.7\textwidth}{X l  l  }
\hline
\hline
Cavity loss rate $\kappa$ & $0.0082$ $\mu$eV&$2.0$ MHz\\
Light-matter coupling $g$ & $0.2050$ $\mu$eV& $50$ MHz\\
Cavity frequency $\omega_0$&  $32.5$ $\mu$eV&$7.86$ GHz  \\ 
Elastic tunneling $t$ & $16.4$ $\mu$eV&$4.0$ GHz\\
Drain tunneling rate $\Gamma_R$ & $16.56$ $\mu$eV&$4.0$ GHz \\ 
Source tunneling rate $\Gamma_L$ & $16.56$ $\mu$eV&$4.0$ GHz \\
Temperature $T$ & $8$ mK& 0.16 GHz \\
\hline
\hline
\end{tabularx}\par
\end{center}
\vspace{10 mm}


\section{Properties of cavity-emitted microwave photon}

In this section, we compute various experimentally-measurable properties 
of the cavity, namely its average photon occupation, emission (power) spectrum, spectral response function, 
as well as the transmission amplitude and phase response of the cavity-emitted microwave photons. It is important to mention that, while performing the phase spectroscopy (transmission amplitude and phase) we relate incoming bosonic modes of the left ($L$) transmission lines with the input microwave signal. While outgoing bosonic modes of the right ($R$) transmission lines construct the output signal \cite{KLH2015}.
For other types of photonic and electronic measurements the ports act as a source for dissipation.
\label{meanPh}

\subsection{Average photon number}
\begin{figure} [pb]
\includegraphics[width=5cm]{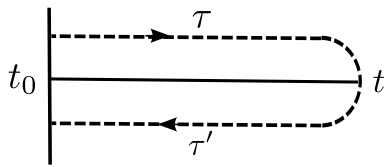}
\caption{The complex-time contour in Keldysh formalism. 
$\tau,\tau'$ are complex-time parameters. 
The contour path starts at some initial time $t_0$, goes to observation time $t$, then comes back to time $t_0$.}
\label{contour}
\end{figure}

\begin{figure} [pb]
\includegraphics[width=12cm]{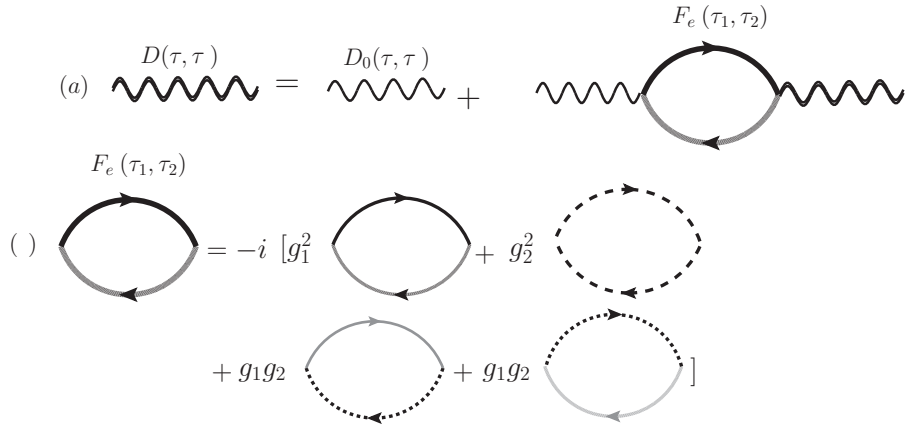}
\caption{(a) Dyson equation for the photon Green's function
$D(\tau,\tau')$ in contour time.
(b) The bubble diagram $F_{el}(\tau_1,\tau_2)$ consists of four dressed electronic Green's functions, represented by  dark solid line ($G^{0}_{11}$), dashed line ($G^{0}_{22}$), dotted line ($G^{0}_{12}$) and grey solid line ($G^{0}_{21}$).
}
\label{Feynman_photon}
\end{figure}

In recent years, the mean photon number became an experimentally accessible quantity\cite{eichler,pettamaserscience,Viennot2013} for QD-cQED setups. We compute the mean photon number $\langle \hat n_c \rangle \equiv \langle \hat a^{\dagger} \hat a \rangle$ in the cavity
using the Keldysh NEGF technique. This method allows us to perform a perturbative expansion (second-order) in the 
electron-photon and cavity-photon bath coupling Hamiltonian, 
while capturing dot-lead interaction effects to all orders (non-perturbative). 
We consider the contour-ordered photon Green's function, 
%
\bea
D(\tau,\tau') &\equiv& -i \langle T_c \hat a(\tau) \hat a^{\dagger}(\tau') \rangle, \nonumber \\
&=& -i \langle T_c \hat a_I(\tau) \hat a_I^{\dagger}(\tau') e^{-i/\hbar \int d\tau_1 \hat H^I_{el-ph}(\tau_1)}\rangle.
\label{dttp}
\eea
It hands over all  components required to calculate various optical signals.
Here, $T_c$ is the contour-ordered operator (see Fig.~\ref{contour}) responsible for the rearrangement of operators according to their contour time. The earlier (later) contour time places operators to the right (left).
In the second line of Eq.~(\ref{dttp}), operators are written in the interaction picture 
with respect to the non-interacting (quadratic) part of the Hamiltonian $\hat H_{el}+\hat H_{ph}$, for which both electronic and photonic Green's functions are known exactly.
The perturbative expansion of Eq.~(\ref{dttp}) generates terms of different orders in the electron-photon coupling $g_{1},g_{2}$.
A naive perturbative calculation with diagrams up to a particular order leads to the
{\it violation} of different symmetry-preserving physical processes,
such as the conservation of charge and energy currents.  
In order to restore basic symmetries, one has to sum over an infinite-subclass of diagrams, taking into account
all electron scattering events  
which are facilitated by the emission or absorption of a single photon quanta $\omega_0$.
This can be achieved by employing the so-called random phase approximation (RPA) \cite{rpa2,rpa1,rpa3}
where a particular type of ring diagrams are summed over,  see Fig.~\ref{Feynman_photon}.
We can represent this infinite summation in a closed Dyson-like (kinetic) equation for $D(\tau,\tau')$,
\bea
D(\tau,\tau')&=& D_0(\tau,\tau')  \nonumber \\
&\!+\!& \int d\tau_1 \int d\tau_2 D_0(\tau,\tau_1) F_{el} (\tau_1,\tau_2) D(\tau_2,\tau').
\eea
$D_0(\tau,\tau')$ is the Green's function of the primary photon which also includes the effect of the secondary photon modes (transmission lines).
$F_{el}(\tau_1,\tau_2) $ corresponds to the bubble diagrams involving the left and right dots' Green's function, see Fig.~\ref{Feynman_photon}. It
describes elastic and inelastic (energy exchange) processes, where electrons in the dots interact with the cavity mode. 
We will later identify the bubble diagrams, in other words, the photon self-energy (connected part), as the density-density correlation 
function of electrons.
In terms of the contour variables, this Green's function can be written as
\bea
F_{el} (\tau_1,\tau_2) &=& -i \, \Bigg[ g_1^2 G^0_{11} (\tau_1,\tau_2) \, {G}^0_{11} (\tau_2,\tau_1)  
+ g_2^2 G^0_{22} (\tau_1,\tau_2) \, {G}^0_{22} (\tau_2,\tau_1) \nonumber \\ &+&
g_1 g_2 \Big\{  G^0_{12} (\tau_1,\tau_2) \, {G}^0_{21} (\tau_2,\tau_1) + {G}^0_{21} (\tau_1,\tau_2) \, G^0_{12} (\tau_2,\tau_1)\Big\} \Bigg], \nonumber \\
&=& -i \,{\rm Tr} \Big[{\bf g}\, {\bf G}^0(\tau_1,\tau_2)\, {\bf g}\, {\bf G}^0(\tau_2,\tau_1) \Big].
\label{eh-prop}
\eea
This function is symmetric under the exchange of the contour time parameters $\tau_1$ and $\tau_2$. 
$G^{0}_{ij} (i,j=1,2)$ are the non-interacting electronic Green's functions (dressed by the arbitrarily strong electron-lead tunnelling Hamiltonian), defined as 
$G^{0}_{ij}(\tau_1,\tau_2)\!=\! - i \,\langle T_c \hat c_{i}(\tau_1) \hat c_{j}^{\dagger}(\tau_2)\rangle$.
The average is performed over the current-carrying steady state, determined by the inverse temperatures $\beta_{L,R}= 1/T_{L,R}$ 
and chemical potentials $\mu_{L,R}$ of the electronic leads. 
Components of the non-interacting electron Green's functions are given in Appendix A. 
In the third line of Eq.~(\ref{eh-prop}) we organize $F_{el}(\tau_1,\tau_2)$ in a matrix form,
 with ${\bf g}$ and ${\bf G}^0$ as $2 \times 2$ matrices with ${\bf g}= {\rm diag}(g_1, g_2)$. 
Expressions for different components of $F_{el}(\tau_1,\tau_2)$ and various relations among them such as the Korringa-Shiba relation are explained in Appendix B. 
It is important to mention that, if the DQDs are further coupled to a phononic environment \cite{pettamaser}, 
${\bf G}^0(\tau_1,\tau_2)$ in the bubble diagrams should be replaced by the interacting ${\bf G}(\tau_1,\tau_2)$, dressed by the phononic interaction (assuming Wick's theorem).

In the steady state limit, different real-time components of $D(\tau,\tau')$ can be obtained. 
The convolution in time domain results in a  multiplicative form in the frequency domain. This gives 
%
\begin{small}
\bea
&&{\bf D}(\omega) =\begin{bmatrix}
                         D^{t}(\omega)& {D}^{<}(\omega)   \\
                          -D^{>}(\omega) & -D^{\bar{t}}(\omega)
\end{bmatrix} =\big[{\bf D}_0^{-1}(\omega)- {\bf F}_{el} (\omega)\big]^{-1} =
\nonumber \\
\!\!\!\!&&
\begin{bmatrix}
                         [D_0^r(\omega)]^{-1}\!-\!\Sigma_{\rm ph}^{<}(\omega)\!-\!F_{el}^t(\omega) \!& \!\!-\Sigma_{\rm ph}^{<}(\omega)\!-\!{F}_{el}^{<}(\omega)   \\
                          \Sigma_{\rm ph}^{>}(\omega)\!+\!{F}_{el}^{>}(\omega) \!&\!\! \!\![D_0^r(\omega)]^{-1}\!+\!\Sigma_{\rm ph}^{>}(\omega)\!+\!F_{el}^{\bar{t}}(\omega)
\end{bmatrix}^{-1}, 
\label{eq:Domega}
\eea
\end{small}
where $t, \bar{t}, <, >$ are the time ordered, anti-time ordered, lesser and greater components of the Green's function.
The primary photon retarded Green's function  is given by
\bea
[D_0^r(\omega)]^{-1}=(\omega-\omega_0)-\Sigma_{ph}^r(\omega),
\eea
with $\Sigma_{\rm ph}^{r,a,<,>}(\omega)$ as different components of the self-energy, materializing due to the coupling of the cavity photon to 
the input and output ports. In the wide-band limit we approximate 
$\Sigma_{\rm ph}^{r}(\omega)= -i \kappa$ (recall that $\kappa= 2 \pi \rho |\nu|^2$ is the decay rate  of the  cavity mode per port). 
We also receive 
$\Sigma^{<}_{\rm ph}(\omega)=-2\,i \, \kappa \,n_{\rm ph}(\omega)$ and $\Sigma^{>}_{\rm ph}(\omega) =-2\,i \,\kappa \,\left[1+n_{\rm ph}(\omega)\right]$ where $n_{\rm ph}(\omega)$ stands for the Bose-Einstein distribution function, evaluated at temperature $T_{ph}$.

We now compute various components of the photon Green's function by inverting the $2\times 2$ matrix in Eq.~(\ref{eq:Domega}). We receive
the greater and lesser components,
\be
D^{</>}(\omega)= \frac{F_{el}^{</>}(\omega) + \Sigma_{ph}^{</>}(\omega)}{\left[\omega\!-\!\omega_0 \!-\!F^{'}_{el}(\omega)\right]^2\!+\! 
\left[F_{el}^{''}(\omega)\!-\!\kappa\right]^2}
\label{less-greater-eq}
\ee
as well as the retarded and advanced Green's functions
\bea
D^{r}(\omega)&=& \Big[\left(\omega\!-\!\omega_0 \!-\!F^{'}_{el}(\omega)\right)\!-\!i \left(F_{el}^{''}(\omega)\!-\!\kappa\right)\Big]^{-1},
\nonumber\\
D^{a}(\omega) &=& \big[D^{r}(\omega)\big]^{*}. 
\label{Dr-eq}
\eea
Here, $F_{el}'(\omega) = Re[F_{el}^r(\omega)] =\left[F_{el}^t(\omega) -F_{el}^{\bar t}(\omega)\right]/2$ and 
$F_{el}''(\omega) = Im[F_{el}^r(\omega)] =\left[F_{el}^{>}(\omega) -F_{el}^{<}(\omega)\right]/2\,i$. In what follows we show that  $F_{el}^{',''}$ play a central role in enhancing gain in the cavity mode. 
For later use we also define the total self-energy $\Pi^{t,{\bar{t}},<,>}(\omega)$ 
which is additive in the electronic and transmission lines induced self-energies, i.e.,
\be
\Pi^{t,\bar{t} <, >}(\omega)= \Sigma_{ph}^{t,\bar{t} <, >}(\omega)+ F_{el}^{t,\bar{t} <, >}(\omega).
\ee
%
%
With this at hand, we identify the mean photon number in the steady-state limit as
\bea
\langle \hat{n}_c \rangle &=& i \int_{-\infty}^{\infty} \frac{d\omega}{2 \pi} D^{<}(\omega)  \nonumber \\
&=& i \int_{-\infty}^{\infty} \frac{d\omega}{2 \pi} \frac{F_{el}^{<}(\omega) + \Sigma_{ph}^{<}(\omega)}
{\left[\omega\!-\!\omega_0 \!-\!F_{el}^{'}(\omega)\right]^2\!+\! \left[F_{el}^{''}(\omega)\!-\!\kappa \right]^2}.
\label{ncint}
\eea
The integration can be performed to include terms
to the second order in the electron-phonon coupling (order $g_i^2$) and in $\kappa$.
We employ the residue theorem to perform the integration.
Upto the second order the poles are located at
$\Big(\omega_0 + F_{el}^{'}(\omega_0) \pm i \big[\kappa\!-\!F_{el}^{''}(\omega_0)\big]\Big)$. 
Assuming $\kappa > F_{el}^{''}(\omega_0)$ the integration in Eq. (\ref{ncint}) then results in 
\be
\int_{-\infty}^{\infty} \frac{d\omega}{2\pi} D^{</>}(\omega)  \approx \frac{F_{el}^{</>}(\omega_0) + \Sigma_{ph}^{</>}(\omega_0)}{2 \, \big[ \kappa \!-\! F_{el}^{''}(\omega_0) \big]}.
\ee
The mean photon number is obtained as
\be
\langle \hat n_c \rangle \!=\!  \frac{F_{el}^{<}(\omega_0) + \Sigma_{ph}^{<}(\omega_0)}{2 \, i \big[ F_{el}^{''}(\omega_0)\!-\!\kappa\big] }\!=\!\frac{\Pi^{<}(\omega_0)} {\Pi^{>}(\omega_0)\!-\! \Pi^{<}(\omega_0)},
\label{eq:nNEGF}
\ee
%
where we alternatively express it in terms of the total self-energy $\Pi^{</>}(\omega_0)$. 
At equilibrium, the metallic leads are maintained at the same chemical potential ($\mu_L\!=\!\mu_R$) and at the same temperature,
equal to the temperature of the photonic environment ($\beta_L\!=\!\beta_R\!=\!\beta_{ph}\!=\!\beta$). 
The detailed balance condition is then satisfied for $F_{el}^{</>}(\omega_0)$ and  $\Sigma_{ph}^{</>}(\omega_0)$,
i.e.,
$F_{el}^{>}(\omega_0) = e^{\beta \omega_0} F_{el}^{<}(\omega_0)$, see Appendix B. 
This ensures the onset  of the
Bose-Einstein distribution for the cavity photon mode  at equilibrium. 

\begin{figure}
\includegraphics[width=17cm]{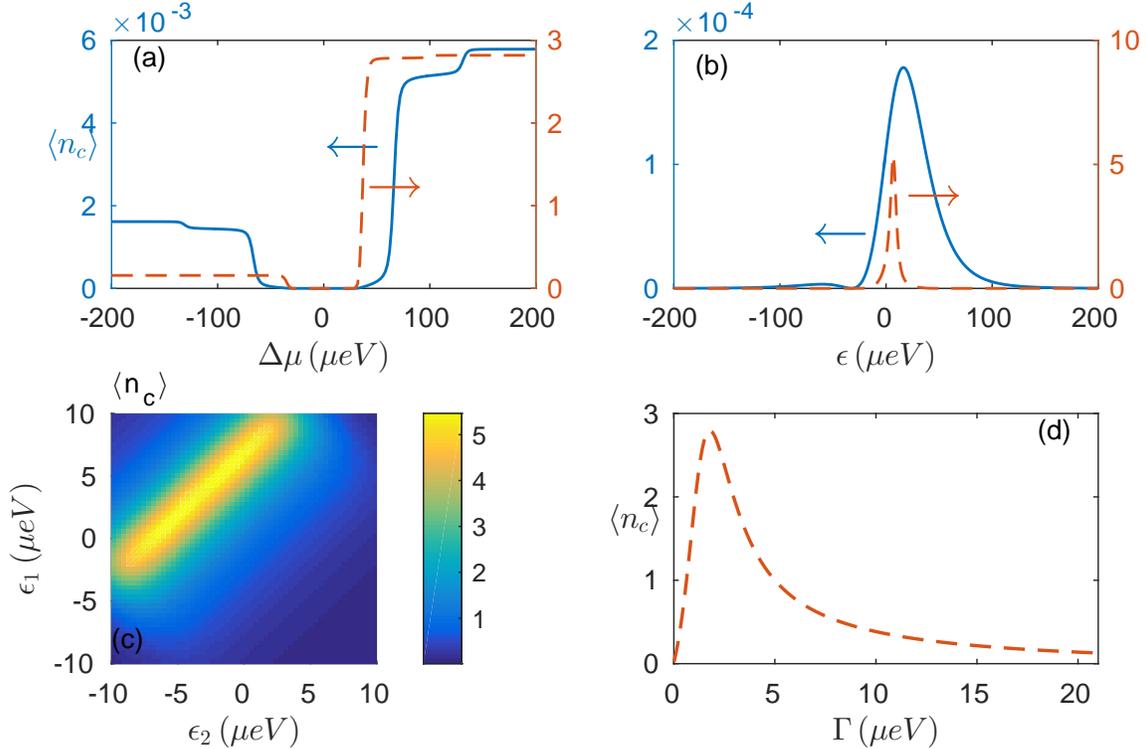}
\caption{(Color online) Average photon number $\langle \hat{n}_c\rangle $ as a function of (a) applied bias voltage $\Delta \mu$, 
(b) energy detuning $\epsilon$, (c) dot energies $\epsilon_1$ and $\epsilon_2$, and (d) dot-lead coupling $\Gamma$. 
Numerical parameters are $g=50\,{\rm MHz}$, $\kappa=0.005\, \mu eV$, $\Gamma=1.656\, \mu eV$, $t=16.4 \,\mu eV$ (dashed), $t=32.8 \,\mu eV$ (solid).
We use $\epsilon=10\, \mu eV$ in panel (a),  $\Delta \mu=50\, \mu eV$ in panel (b), 
and $t=16.4 \, \mu eV, \Delta \mu =50\, \mu eV$ in panels (c)-(d). Other parameters are the same as in the Table.} 
\label{navg}
\end{figure}

From Eq.~(\ref{eq:nNEGF}) we can further obtain the mean-square displacement of the cavity mode as 
\bea
\langle \hat X^2 \rangle =  1 + 2 \langle \hat{n}_c \rangle =\frac{\Pi^{>}(\omega_0) \!+\! \Pi^{<}(\omega_0) } {\Pi^{>}(\omega_0) \!-\! \Pi^{<}(\omega_0)}.
\eea
%


In Fig.~\ref{navg} we show the average cavity occupation number $\langle \hat{n}_c \rangle$ as a function of applied bias, 
energy detuning and dot-lead coupling. Unless otherwise stated we define here and below the detuning $\epsilon= \epsilon_1\!-\!\epsilon_2$ and enforce $\epsilon_1\!=\!-\epsilon_2\!=\! \epsilon/2$, with matter-light coupling $g\!=\!g_1\!=\!-g_2$. For the spectral functions of the metallic leads we make the wide-band approximation,
 and fix $\Gamma_L\!=\!\Gamma_R\!=\!\Gamma$. We also set the equilibrium Fermi energy of the metal leads at zero and change the bias symmetrically with $\mu_L\!=-\mu_R\!=\! \Delta \mu /2$. The temperature of the two metals and the transmission lines are chosen to be identical.
 
In Fig. \ref{navg}(a) we study the average photon occupation as a function of bias voltage. 
We find that $\langle \hat{n}_c\rangle$ increases in two steps. At the first step, $\Delta \mu > \omega_0$, 
tunnelling electrons acquire sufficient energy to interact with the cavity mode and generate photons. 
The second step arises due to the additional resonance situation at $\Delta \mu \sim \sqrt{\epsilon^2 + 4 t^2} + 2 \, \omega_0$. 
At this bias electrons arriving from the left metal at $\Delta \mu/2$ deposit energy ($\omega_0$) to the cavity mode, 
allowing them  to resonantly cross the junction. 
In the positive detuning case $(\epsilon>0)$, examined here, $\langle \hat{n}_c\rangle$ saturates at lower values for reverse (negative) bias, in comparison to that in the forward (positive) bias. This cavity-number asymmetry with respect to bias reflects the structural asymmetry in the DQD system.  
With increasing tunnelling strength $t$, fast interdot charge transfer results in effectively weak interaction between electrons and the cavity mode, therefore showing low values for the average photon number.

In Fig. \ref{navg}(b) we plot $\langle \hat{n}_c \rangle$ as a function of detuning and observe a significant enhancement in the photon number. 
For the bare system Hamiltonian, strong photon emission into the cavity is expected when $\omega_0 = \sqrt{\epsilon^2 + 4 t^2}$,
satisfying energy conservation. 
The position of the peaks in Fig. \ref{navg}(b) is renormalized with respect to bare values due to dot-lead coupling. 
Close to the resonance condition a sharp increase in the photon number is observed, potentially creating a lasing function. 
For large detuning the DQD system does not well interact with the cavity mode, resulting in a vanishing photon occupation. 
To elucidate this behavior further we display $\langle \hat{n}_c \rangle$ as a function of dot energy levels $\epsilon_1$ and $\epsilon_2$ 
in Fig. \ref{navg}(c). We see similar trends with large photon generation ($\langle \hat{n}_c\rangle \sim 5$) for $\epsilon_1 > \epsilon_2$. 

The nonlinear and non-monotonic behavior of $\langle \hat{n}_c \rangle$ as a function of dot-lead coupling $\Gamma$ is demonstrated in Fig. \ref{navg}(d).
At weak coupling and finite detuning,  $\langle \hat{n}_c\rangle$ increases with $\Gamma$, as the photon number in the cavity is amplified 
by charge transfer through the DQD system. In contrast, when the dot-leads coupling is strong, the renormalization 
and broadening of the dot energy levels allow electrons to tunnel through the DQD on a short timescale, 
only briefly interacting with the cavity photons, thus resulting in limited photon generation.

\begin{figure}
\includegraphics[width=15cm]{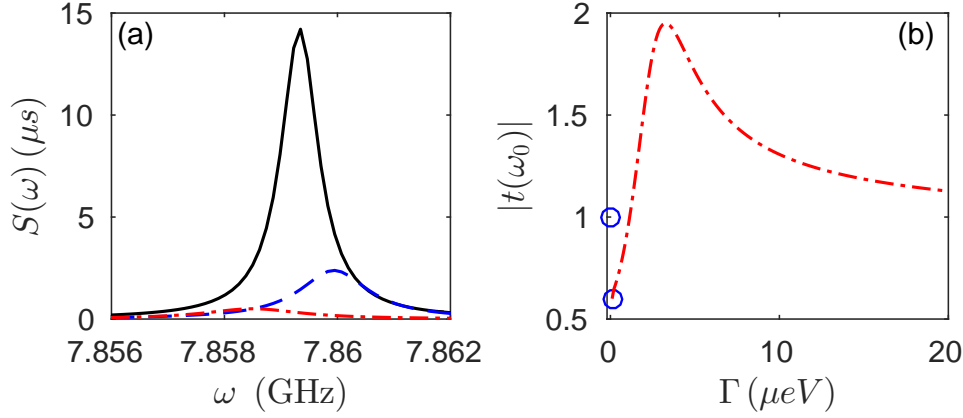}
\caption{(Color online) (a) Power spectrum $S(\omega)$ as a function of frequency for different dot-lead coupling. $\Gamma\!=\!0.056 \mu eV$ (dashed), $1.656\, \mu eV$ (solid), $16.56\, \mu eV$ (dashed-dotted).
(b) Absolute value of transmission as a function of dot-lead coupling $\Gamma$ for a fixed bias.
Numerical parameters are $g=50\,{\rm MHz}$, $\kappa=0.005\, \mu eV$, $t=16.4 \,\mu eV$, $\epsilon=10\, \mu eV$,  $\Delta \mu=500\, \mu eV$. Other parameters are the same as in the Table. Circles represent the values of $|t(\omega_0)|$ at $\Gamma\!=\!0$ and at finite but very weak $\Gamma$.} 
\label{Sw}
\end{figure}

\subsection{Power spectrum and spectral function for the cavity mode}
Next we look at the power spectrum and the spectral function of the cavity mode. The power spectrum has recently been measured for a similar setup \cite{pettamaserscience}. In the stationary limit, we can immediately obtain the emission (power) spectrum in terms of the lesser Green's function 
\bea
S(\omega) &=& \int_{-\infty}^{\infty} dt \, \langle \hat{a}^{\dagger}(0) \hat{a}(t) \rangle e^{i \omega t} = i\, D^{<}(\omega) \nonumber \\
&=&i \frac{F_{el}^{<}(\omega) + \Sigma_{ph}^{<}(\omega)}{\big[\omega\!-\!\omega_0 \!-\!F^{'}_{el}(\omega)\big]^2\!+\! \big[F_{el}^{''}(\omega)\!-\!\kappa\big]^2}.
\eea
It is obvious from this definition that $\langle \hat n_c \rangle =  \int \frac{d\omega}{2\pi} S(\omega)$.  
At the cavity frequency $\omega_0$, $S(\omega)$ is given by
\be
S(\omega_0)= -\frac{ 2 \,\langle \hat{n}_c \rangle \, \big[F_{el}''(\omega_0) -\kappa\big] }{\big[F_{el}'(\omega_0)\big]^2 + \big[F_{el}''(\omega_0) -\kappa\big]^2}.
\ee
Similarly, we obtain an expression for the spectral response function of the resonator, 
defined as the difference between the retarded and advanced photon Green's function  
\bea
A_{ph}(\omega)&=& i \left[ D^r(\omega)-D^a(\omega)\right] \nonumber \\
&=& \frac{-2\, \big[F_{el}^{''}(\omega)\!-\!\kappa\big]}{\big[\omega\!-\!\omega_0\!-\!F_{el}^{'}(\omega)\big]^2 + \big[F_{el}^{''}(\omega)\!-\! \kappa\big]^2},
\eea
with the normalization condition (sum rule) $\int \frac{d\omega}{2 \pi} A_{ph}(\omega)\!=\! \langle [\hat{a}, \hat{a}^{\dagger}]\rangle=1$. This can be proved as well from the above equation invoking the residue theorem as explained before.
In the absence of the light-matter interaction the cavity spectral function $A_{cav}(\omega) = \frac{2 \kappa}{(\omega-\omega_0)^2 + \kappa^2}$ trivially satisfies the sum rule. The amplitude of the spectral function at $\omega_0$ is related to the emission spectrum as $A_{ph}(\omega_0)= S(\omega_0)/\langle \hat{n}_c \rangle$. In Fig.~\ref{Sw}(a) we plot the power spectrum $S(\omega)$ for different dot-lead coupling. It shows a nonmonotonic behavior with respect to tunnelling strength, with the  maximum value taking place at an intermediate value for the tunnelling. The brodening, which is of the order of several MHz (1-3 MHz), results from the interplay between $F_{el}''$ and $\kappa$, the two different sources of dissipation.


\subsection{Phase spectroscopy: Transmission and Phase}
We calculate the transmission amplitude and the phase response of the emitted microwave photons by following the input-output relations 
in Refs.~\onlinecite{input_output_Clerk,simon16, marco14}. Recall that for phase spectroscopy measurements (performed via heterodyne detection \cite{petta2014})the bosonic modes of the left ($L$) and the right ($R$) transmission lines are related with the input and output microwave signal, respectively \cite{KLH2015}. 
The transmission function $t(\omega)$ reads as \cite{input_output_Clerk,simon16, marco14}
\be
t(\omega) \!=\! \frac{i \kappa}{(\omega-\omega_0) + i \kappa -\chi_{el}^r(\omega)} \!=\!i \, \kappa \,D^r(\omega),
\label{eq:tdef}
\ee
where $\chi_{el}^r(\omega)$ can be shown to be proportional to the electronic charge susceptibility as 
$\chi_{el}^{r}(t\!-\!t')= \sum_{\alpha , \beta \in 1,2} g_{\alpha} \, g_{\beta}\, \Lambda^{el}_{\alpha \beta}(t-t')$ \cite{simon16, marco14}.
Here 
\be
\Lambda^{el}_{\alpha \beta}(t-t')= -i \theta(t-t') \, \big\langle \big[\hat{n}_{\alpha}(t), \hat{n}_{\beta}(t')\big]\big\rangle_{el}  
\ee
stands for the electron density response function, $\alpha,\beta=1,2$ are indices for the dots,
 $\langle \cdots \rangle_{el}$ refers to the average over the electronic degrees of freedom in the nonequilibrium steady state. 
In our formulation, following Eq.~(\ref{Dr-eq}) we identify the transmission to be proportional to the retarded Green's function $D^r(\omega)$ of the cavity photon mode, which in the time-domain is precisely the photon response function, given by   
\be
D^r(t-t')=-i \,\theta(t-t') \langle [\hat{a}(t), \hat{a}^{\dagger}(t')]\rangle_T. 
\ee
Therefore, we note that $\chi_{el}^r(\omega)$ stands for the retarded component of the bubble diagram $F_{el}^r(\omega)$. 
Here $\langle \cdots \rangle_T$ represents average over the combined photonic-electronic steady state density operator.

\begin{figure}
\includegraphics[width=15cm]{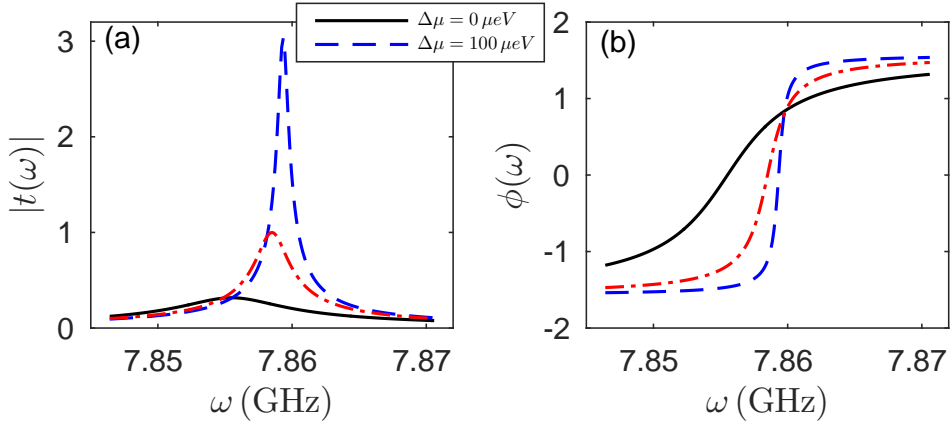}
\caption{(Color online) Spectroscopy of cavity-emitted photons.
(a) Transmission (absolute value) and (b) phase response as a function of incoming photons frequency $\omega$. 
Parameters are $\kappa=0.005 \, \mu eV$, $\Gamma=1.656 \, \mu eV$, $g \!=\! 50 \,{\rm  MHz}$, $\epsilon=10 \, \mu eV$, 
$\Delta \mu = 0\, \mu eV$ (solid) and $\Delta \mu = 100\, \mu eV$ (dashed). 
The red dashed-dotted line was obtained at $g\!=\!0$. 
Other parameters are same as in the Table. 
The sum rule for transmission is $\int \frac{d\omega}{2\pi} \,t(\omega) = \kappa/2$.} 
\label{trans_omega}
\end{figure}

We further write $t(\omega)= |t(\omega)| e^{i \phi(\omega)}$ and identify the real and imaginary parts of the transmission function,
\bea
{Re}\,[t(\omega)]&=& -\kappa \,{Im}\, [D^r(\omega)] =\kappa \frac{\kappa \!-\!F_{el}''(\omega)}{\big[\omega\!-\!\omega_0\!-\!F_{el}{'}(\omega)\big]^2 + \big[F_{el}''(\omega)\!-\!\kappa\big]^2}, \nonumber \\
{Im}\,[t(\omega)]&=& \kappa \,{Re}\, [D^r(\omega)] = \kappa \frac{ \omega\!-\!\omega_0\!-\! F_{el}'(\omega)}{\big[\omega\!-\!\omega_0\!-\!F^{'}_{el}(\omega)\big]^2 + \big[F_{el}''(\omega)\!-\!\kappa\big]^2},
\eea
with the sum rule $ \int_{-\infty}^{\infty} \frac{d\omega}{2\pi} \,t(\omega)= \kappa/2$. 
Specifically, we get  $\int_{-\infty}^{\infty} \frac{d\omega}{2\pi} \,Re[t(\omega)]= \kappa/2$ and $\int_{-\infty}^{\infty} \frac{d\omega}{2\pi} \,Im[t(\omega)]= 0$. 
We further note that the real part of the transmission function provides a direct measure for the spectral function of the cavity photon. 
The phase of emitted photons is
\be
\tan \phi(\omega) = -\frac{{Re}\, [D^r(\omega)]}{{Im}\, [D^r(\omega)]}= -\frac{\omega-\omega_0-F_{el}'(\omega)}{F_{el}^{''}(\omega)-\kappa}.
\ee
Since we are mainly interested in the absolute value of the transmission function 
and the value of the phase response at the frequency of the cavity mode $(\omega=\omega_0)$, we evaluate
\bea
|t(\omega_0)| &=& 
\frac{\kappa}{\Big[\big(F_{el}^{'}(\omega_0)\big)^2 \!+\! \big(F_{el}^{''}(\omega_0)\!-\!\kappa\big)^2\Big]^{1/2}}, \nonumber \\
\tan \phi(\omega_0) &=& 
\frac{F_{el}^{'}(\omega_0)}{F_{el}^{''}(\omega_0)\!-\!\kappa}. 
\eea
%
Note that both the real and imaginary parts of $F_{el}(\omega)$ show nontrivial dependence on bias voltage through the Keldysh component 
of the electronic Green's function $G_0^{k}(\omega)$, see Appendix B.


In Fig.~\ref{trans_omega} we plot the absolute value of the transmission and the phase response as a function of the incoming photon frequency $\omega$ under different bias voltages $\Delta \mu$. When the cavity is decoupled from the DQD, $g=0$ (dashed-dotted line), the transmission reaches unity at 
$\omega=\omega_0$, and the broadening is determined by $\kappa$.  As well, 
the phase response is zero at the resonant frequency, and it approaches $\pm \pi/2$ in the off-resonant regime. 
For finite $g$---yet at zero bias---charge fluctuations in the dots introduce shift in the transmission peak, further reducing the maximum amplitude. 
The frequency shift depends on the real part of the charge susceptibility $F_{el}'(\omega)$, whereas the broadening 
reflects the difference between $F_{el}''(\omega)$ and $\kappa$. The phase response is zero when the transmission
is at maximum.
Most interestingly, we find that the absolute value of the transmission coefficient can be greatly enhanced---beyond unity---at finite bias, once  $F^{''}_{el}>0$. This situation is elaborated below.



\begin{figure}
\includegraphics[width=15cm]{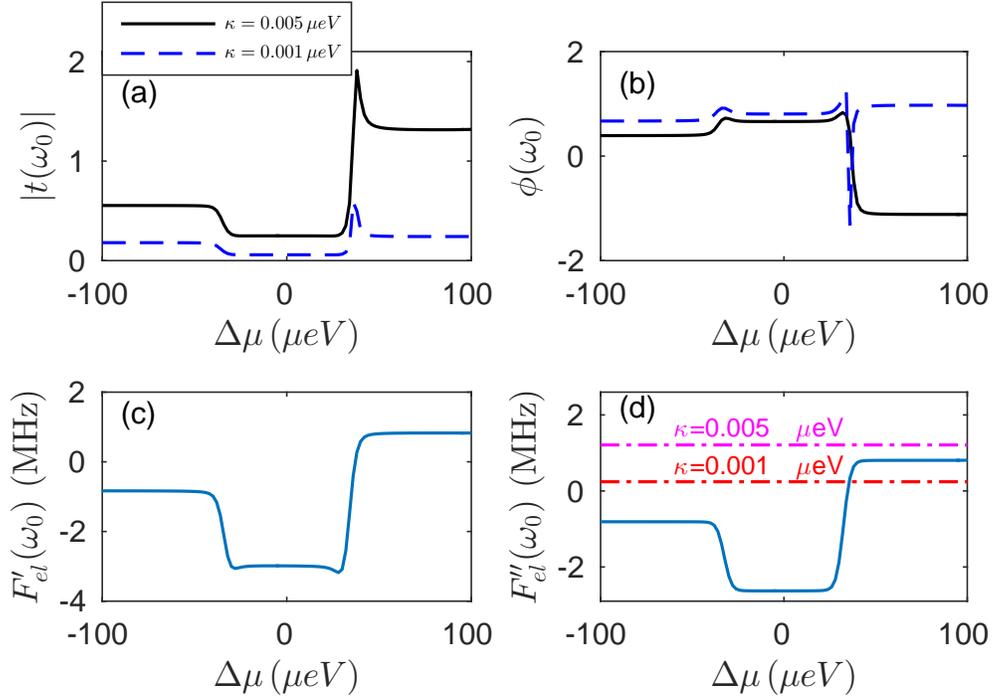}
\caption{(Color online) Controlling transmitted photons by a DC bias voltage across the dots.
(a) Transmission and (b) phase, as a  function of bias voltage $\Delta \mu$ at the bare cavity frequency $\omega_0$.
In panels (c) and (d) we display the real and imaginary parts of the bubble electronic Green's function $F_{el}(\omega_0)$, respectively,
illustrating that a jump in transmission (and a dip in phase) precisely occurs once 
$\kappa= F_{el}^{''}$ (panel d).
Parameters are $\kappa=0.005 \, \mu eV$ or 1.21 MHz (solid), $\kappa=0.001 \, \mu eV$ or 0.24 MHz (dashed), $\epsilon=10\,\mu eV$, 
$\Gamma=1.656 \, \mu eV$, $g \!=\! 50 \,{\rm  MHz}$, $t=16.4\, \mu eV$. 
Other parameters are given in the Table.} 
\label{trans_phase_bias}
\end{figure}

Fig.~\ref{trans_phase_bias} displays one of the central results of our work: The transmitted photon signal can be 
significantly enhanced at {\it finite} bias voltage, 
once the electronic system is fine-tuned to counteract dissipation from the (photonic) transmission lines.
We study the behavior of the transmission coefficient and phase response at the bare
cavity resonance frequency as a function of bias, for different cavity decay rates $\kappa$.
We find that the transmission, or photon gain, increases for $|\Delta \mu| > \omega_0$,
 and it saturates at high (positive and negative) biases. 
This behavior agrees  with our observations for $\langle \hat{n}_c \rangle$ in Fig.~\ref{navg}(a).
However, at a certain finite voltage (here around 30$\mu eV$), the transmission jumps above the
asymptotic value (panel a), while the phase response shows a sudden dip (dashed line in panel b). 
This sudden jump takes place precisely as
the electronic part of the system acts to cancel out relaxation effects due to the tranmission lines, 
$F_{el}^{''}(\omega_0)= \kappa$, see panel (d).

%
%

\begin{figure}
\includegraphics[width=15cm]{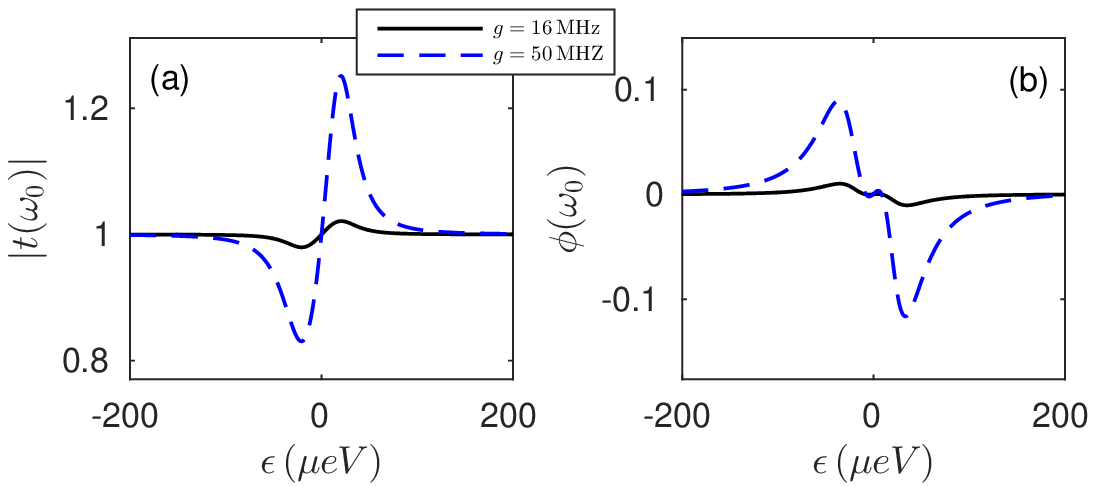}\\
\includegraphics[width=15cm]{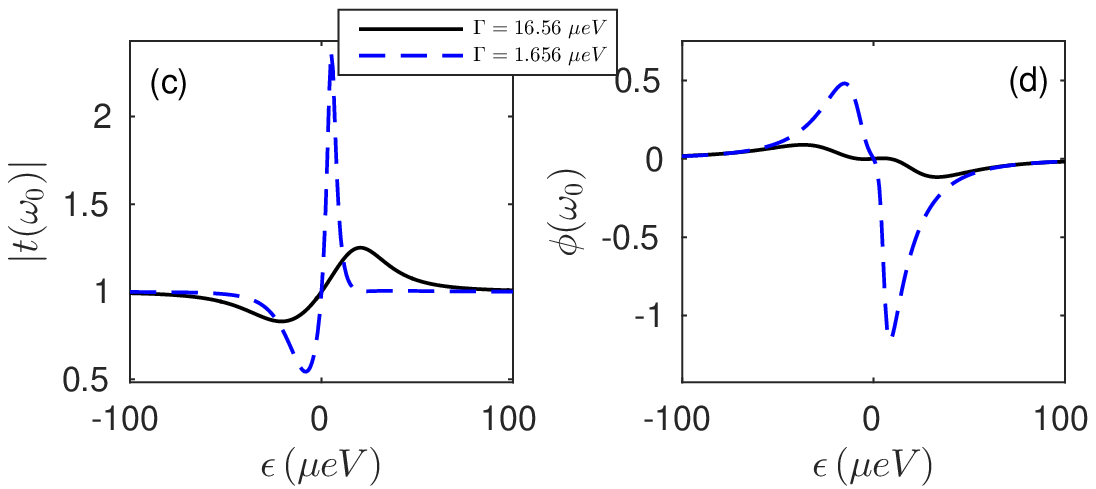}
\caption{(Color online) 
Effect of detuning the DQD levels on the cavity signal.
(a) and (c) Absolute value of transmission coefficient $|t(\omega_0)|$, and (b) and (d)
phase response $\phi(\omega_0)$ of the transmitted signal, displayed as a function of 
detuning $\epsilon$ at high bias using $t\!=\!16.4\, \mu eV$, $\omega_0= 32.5\, \mu eV$, 
cavity loss rate $\kappa=0.005\, \mu eV$, $\Delta \mu= 250 \mu eV$. 
$\Gamma=16.56 \, \mu eV$ in panels (a)-(b), $g \!=\! 50 \,{\rm  MHz}$ in panels (c)-(d). 
Other parameters are same as in the Table.}
\label{trans_phase_detu}
\end{figure}

We turn to the high bias limit and study in Fig.~\ref{trans_phase_detu}
the role of energy detuning $\epsilon$ on the transmission and phase response.
In particular, we examine the dependence of $|t(\omega_0)|$ and $\phi(\omega_0)$ on the electron-photon coupling strength $g$ and on 
the incoherent tunnelling rate $\Gamma\!=\!\Gamma_L\!=\!\Gamma_R$. 
Panels (a) and (c) display the transmission amplitude, showing gain $|t|>1$ (dip, $|t|<1$) on the positive (negative) side of detuning $\epsilon > 0\, 
(\epsilon < 0)$ as a result of coupled electron photon transport processes: 
For positive detuning and positive bias, electron transport through the DQD system proceeds via inter-dot tunnelling, thereby reducing 
the energy of electrons via photon emission. 
In contrast, for negative detuning electron transport proceeds assisted by photon absorption, reflected by the dip in the transmission coefficient. 
This gain mechanism can be also corroborated with the amplification of $\langle \hat{n}_c \rangle$. 
Finally, note that at zero detuning ($\epsilon=0$), direct elastic tunnelling dominates over photon-induced contributions. For very large detuning, marginal 
charge flow through the dots results in an effectively minuscule electron-photon interaction. 
These two limits lead to unit transmission amplitude and zero phase response. 

The relative strengths of $F_{el}^{''}$ and $\kappa$ determines the gain and loss values in the transmission amplitude. 
Plots of the real and imaginary components of $F_{el}$, displayed as a function of $\epsilon$, are included in Appendix B (Fig. \ref{self-energy}). 
Maximum gain is achieved when $F_{el}^{''}=\kappa$. 
Since $F_{el}^{''} \propto g^2$, and 
typically $F_{el}^{''} <\kappa$, 
increasing $g$ shows a significant enhancement in gain and similarly loss. 

\begin{figure}
\includegraphics[width=18cm]{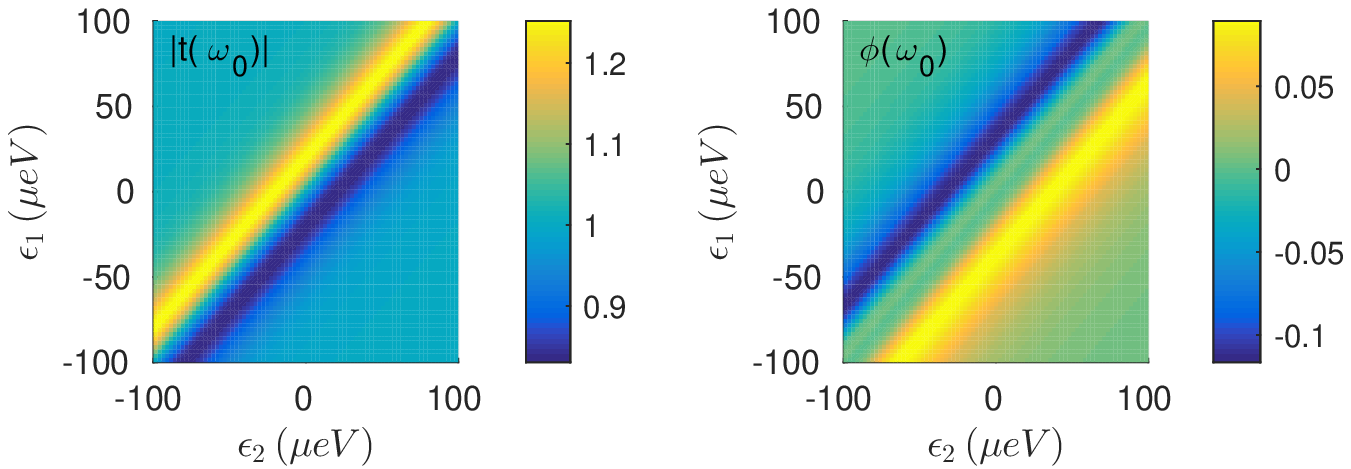}
\caption{(Color online) Control of cavity response by tuning the DQD site energies.
Plot of
(a) absolute value of transmission coefficient $|t(\omega_0)|$ and (b) phase response $\phi(\omega_0)$ 
of the transmitted signal as a function of $\epsilon_1$ and $\epsilon_2$ at high bias $\Delta \mu= 250\, \mu eV$. 
Here $\Gamma=16.56 \, \mu eV$, $g \!=\! 50 \,{\rm  MHz}$. Other parameters are the same as in the Table.}
\label{trans_phase_detu_contour}
\end{figure}

The dependence of $|t(\omega_0)|$ and $\phi(\omega_0)$ on the system-lead coupling strength 
is examined in Fig.~\ref{Sw}(b), showing a nonliner behaviour, and in Figs.~\ref{trans_phase_detu}(c) and (d). 
The transmission is high in the sequential tunnelling regime (intermediate dot-lead coupling),
whereas for large coupling renormalization and broadening of peaks lead to reduced gain.
The reason is that the dwelling time of electrons in the dots is long $(\sim 1/\Gamma$) at weak coupling, 
realizing an effectively significant electron-photon interaction. Indeed, 
figure \ref{self-energy} in Appendix B demonstrates  that
small $\Gamma$ returns large values for $F_{el}^{{'},{''}}$, 
thus a significant enhancement in $|t(\omega_0)|$ as a whole. 
Upon increasing the coupling strength $\Gamma$, the dots energies become broadened, thus electrons flow across the device 
without interacting with the cavity mode. This scenario shows small gain and loss. At very weak coupling, $F_{el}^{''}(\omega_0) \ll \kappa$, the electronic medium introduces only dissipation, responsible for the sharp drop in transmission, see Fig.~\ref{Sw}(b)

Plots of $|t|$ and $\phi$, at the bare cavity frequency, as a function of $\epsilon_1$ and $\epsilon_2$, reveal that 
degenerate quantum dots do not influence the cavity, 
with $|t(\omega_0)|=1$ and $\phi(\omega_0)=0$, see Fig.~\ref{trans_phase_detu_contour}.
In contrast, at positive (negative) detuning, approximately satisfying $\sqrt{4t^2-\omega_0^2} \sim \pm(\epsilon_1-\epsilon_2)$, gain (loss) is observed.

\begin{figure}
\includegraphics[width=15cm]{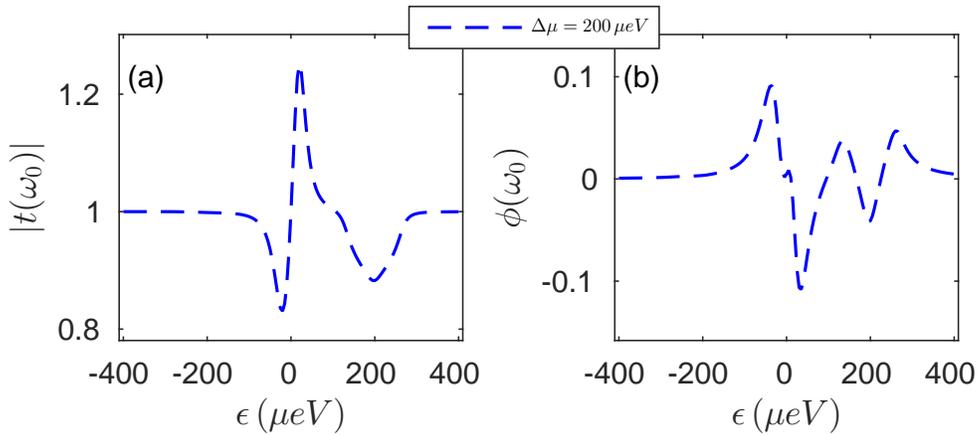}
\caption{(Color online) Signatures of finite voltage bias on (a) $|t(\omega_0)|$ and (b) $\phi(\omega_0)$,
studied as a function of $\epsilon$. 
Here, $\Delta \mu =200 \,\mu eV$, $\Gamma=16.56 \, \mu eV$, $g \!=\! 50 \,{\rm  MHz}$. 
Other parameters are the same as in Fig.~\ref{trans_phase_detu}.}
\label{finite-bias}
\end{figure}

In Fig.~\ref{finite-bias} we study the effect of a finite bias voltage, with energy detuning lying within the voltage bias window, on the transmission 
coefficient and phase response. 
Comparing finite-intermediate voltage results to Fig. \ref{trans_phase_detu}, where high bias was employed, we note here an additional dip at $\epsilon \sim \Delta \mu$,
reflecting photon-assisted charge transfer processes from the right dot to the left dot. Correspondingly, a jump in phase is detected
at the same value of detuning.

\subsection{Special limits, scaling and universality}

We discuss here scaling relations for measures related to the  cavity, in different parameter regimes.
We begin by considering the high bias and zero temperature limit. 
We further assume that the dot-lead coupling is strong, larger than detuning, $\Gamma>\epsilon$. 
This situation is experimentally relevant and thereby potentially testable. We use Eq.~(\ref{bubble-GF}) along with the expressions for the non-interacting electronic Green's functions (see Appendix A). In this limit we found that 
\bea
F_{el}^{<}(\omega_0)&=& -i \,\frac{4 g^2 t^2 \omega_0}{\pi\,\tilde{\Gamma}^4} \Big(\frac{\Delta \mu}{\omega_0}-1\Big), 
\nonumber \\
F_{el}^{>}(\omega_0) &=& F_{el}^{<}(-\omega_0),
\eea
with $\tilde{\Gamma}={\Gamma}/2$. Further assuming that  $\kappa  \ll F_{el}^{</>}$, the average photon number in Eq.~(\ref{eq:nNEGF}) reduces to 
\be
\langle \hat{n}_c \rangle = \frac{1}{2}\Big(\frac{\Delta \mu}{\omega_0}\!-\!1\Big), 
\ee
with an effective temperature  $T_{eff}= \Delta \mu/2$. 
It is remarkable to note that this linear scaling, $\langle \hat{n} \rangle  \propto \Delta \mu$, 
is universal to the \textit{Holstein-like} class of models \cite{Dima_holstein, Utsumi_holstein,Jinshuang_holstein, Bijay_recon}.  
In the opposite limit $\kappa \gg F_{el}^{</>}$ (though keeping $\Gamma>\epsilon$),
the electronic part effectively decouples from the cavity. 
As a result, the cavity equilibrates with the secondary photon bath (ports),
$\langle \hat{n}_c \rangle=n_{\rm ph}(\omega_0)$, and the transmission amplitude goes to unity.


Another interesting limit is the large detuning $\epsilon\gg \Gamma$ and high bias case, where we obtain  
\bea
F_{el}^{<}(\omega_0)&=& -i \,\frac{64 g^2 t^2 \Gamma^2 \omega_0}{\pi\,\epsilon^6} \Big(\frac{\Delta \mu}{\omega_0}-1\Big), \nonumber \\
F_{el}^{>}(\omega_0) &=& F_{el}^{<}(-\omega_0). 
\eea
In this case, $F_{el}^{</>}(\omega_0) \propto 1/\epsilon^6$, the electronic current is negligible, 
and electron-photon coupling is effectively small. 
Again we find that when $\kappa\ll F_{el}^{</>}$,
$\langle \hat{n}_c \rangle = \frac{1}{2}\Big(\frac{\Delta \mu}{\omega_0}\!-\!1\Big)$,
and in the opposite limit,  $\kappa\gg F_{el}^{</>}$, the cavity occupation number is thermal.

\section{Electronic properties: Steady state charge current}
\label{CGF-NEGF}
We study the electronic properties of the DQD system, 
focusing on the steady state charge current at the left contact. It is given by the powerful Meir-Wingreen formula \cite{mwprl, Jauho}, 
valid for an arbitrary large light-matter interaction and dot-lead coupling
\be
I_L = e \int_{-\infty}^{\infty} \frac{d \omega}{2 \pi} {\rm Tr} \Big[ {\bf G}^{>}(\omega) {\bf \Sigma}_L^{<}(\omega)\!-\! {\bf G}^{<}(\omega) {\bf \Sigma}_L^{>}(\omega) \Big].
\label{MW}
\ee
%
\begin{figure} [pb]
\includegraphics[width=8cm]{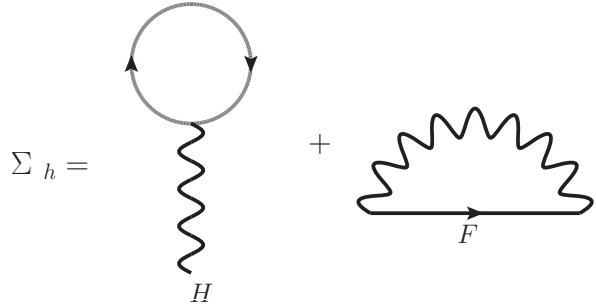}
\caption{Hartree ($H$) and Fock ($F$) diagrams for electrons in contour time. For expressions see Eq.~(\ref{H-F})}.
\label{Hartree}
\end{figure}
Here ${\bf G}^{</>}(\omega)$ are the lesser and greater components of the electronic Green's function, 
fully dressed by the leads and the  electron-photon interaction. 
As before, we obtain these components by relying on the Dyson equation, performing second order perturbation expansion
in the electron-photon interaction \cite{jeremy, Wang_Jian}. We write  
\bea
{\bf G}(\tau,\tau')&=& {\bf G_0}(\tau,\tau')  \nonumber \\
&\!+\!& \int d\tau_1 \int d\tau_2 {\bf G_0}(\tau,\tau_1) {\bf \Sigma}_{ph}(\tau_1,\tau_2) {\bf G}(\tau_2,\tau'),
\eea
with ${\bf \Sigma}_{ph}(\tau_1,\tau_2)$ as the nonlinear electronic self-energy arising due to the photonic degrees of freedom. 
Calculating it  up to $O(g^2)$ we receive the Hartree (H) and Fock (F) terms \cite{galperinprb, Misha_IETS} (see Fig.~\ref{Hartree}), 
\bea
\big[\Sigma_{ph}^{H}\big]_{jk}(\tau_1,\tau_2)&=& -i \,\delta(\tau_1,\tau_2) \,\delta_{jk} \,g_j \,\int d\tau_3 {\rm Tr} \Big[{\bf{g}} \, {\bf G}_0(\tau_3,\tau_3) \Big] D_0(\tau_1,\tau_3), \nonumber \\
\big[\Sigma_{ph}^{F}\big]_{jk}(\tau_1,\tau_2)&=& i\, g_j\, g_k \, D_0(\tau_1,\tau_2) \, \big[{\bf G}_{0}\big]_{jk}(\tau_1,\tau_2),
\label{H-F}
\eea
with ${\bf \Sigma}_{ph}(\tau_1,\tau_2)= {\bf \Sigma}_{ph}^{H}(\tau_1,\tau_2) + {\bf \Sigma}_{ph}^{F}(\tau_1,\tau_2)$.
Following the Keldysh equation, we gather the lesser and greater components in the frequency domain as 
${\bf G}^{</>}(\omega)= {\bf G}^{r}(\omega) {\bf \Sigma}^{</>}_{tot}(\omega) {\bf G}^{a}
(\omega)$, with the total self-energy as the sum of left and right-lead self-energies, as well as the nonlinear component, 
${\bf \Sigma}^{</>}_{tot}(\omega) = {\bf \Sigma}^{</>}_L(\omega)+ {\bf \Sigma}^{</>}_R(\omega) + {\bf \Sigma}_{ph}^{</>}(\omega)$.
%
 \begin{figure}
\includegraphics[width=15cm]{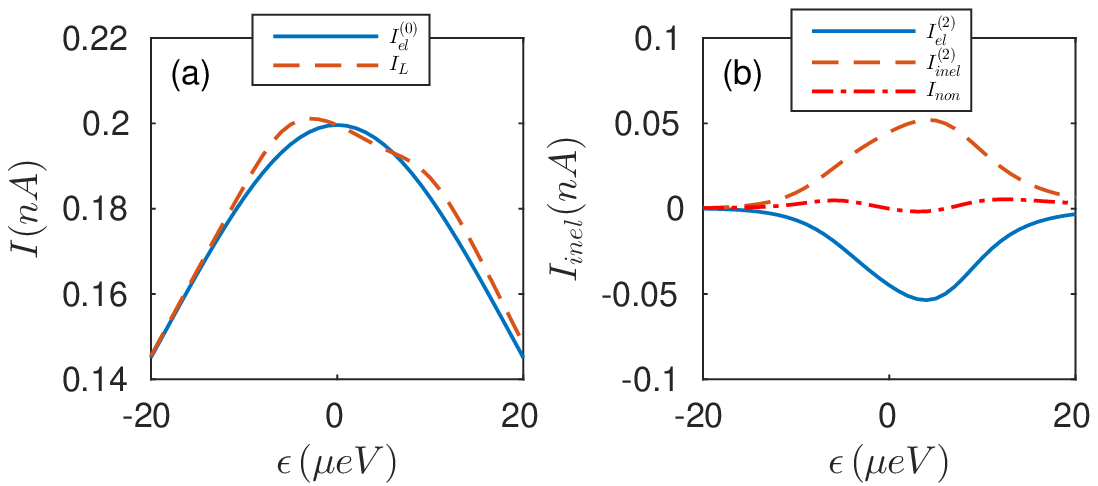}\\
\includegraphics[width=15cm]{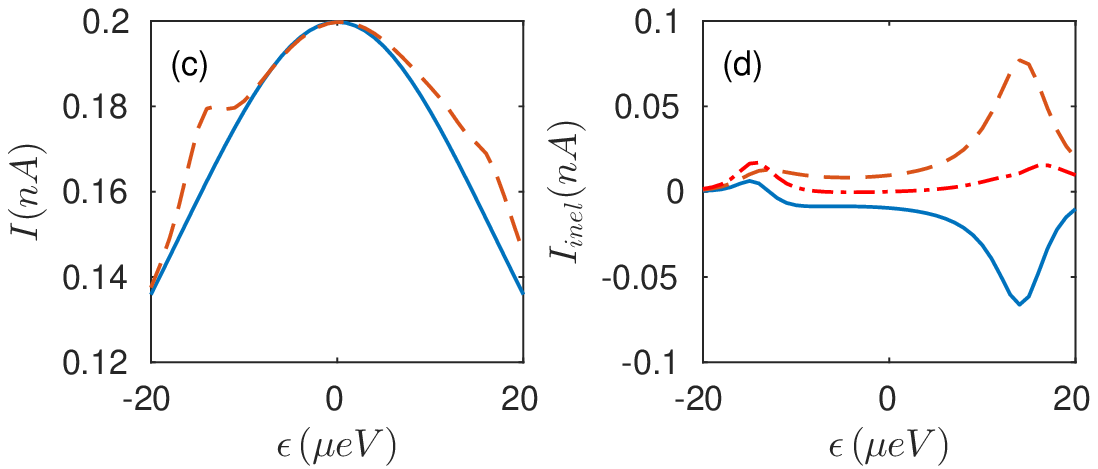}
\caption{(Color online) (a),  (c) Charge current as a function of detuning $\epsilon$. 
The nonlinear contribution to the current  $I_{non}=I_{el}^{(2)}+I_{inel}^{(2)}$ is shown separately in panels (b) and (d).
Parameters are $\kappa=0.005\, \mu eV$,  
$g=150\, {\rm MHz}$, $\Gamma=1.656 \,\mu eV$, $\Delta \mu =60\, \mu eV$, with $t= 16.4 \,\mu eV$ in panels (a) and (b),  $t= 14.6 \,\mu eV$
in panels (c)-(d).} 
\label{current_fig}
\end{figure}
%
Substituting these expressions into the Meir-Wingreen formula we organize the charge current formula, 
written as a sum of elastic and inelastic contributions i.e., $I_L=I_{el} + I_{inel}$, with 
\bea
I_{el} = e \int_{-\infty}^{\infty} \frac{d\omega}{2 \pi} {\rm Tr} \Big[ {\bf G}^r(\omega) {\bf \Gamma}_L(\omega) {\bf G}^a(\omega) {\bf \Gamma}_R(\omega) \Big] \big[f_L(\omega)-f_R(\omega)\big],
\eea
and
\bea
I_{inel}= e \int_{-\infty}^{\infty} \frac{d\omega}{2 \pi} {\rm Tr} \Big[{\bf G}^r(\omega) {\bf \Sigma}_{ph}^{>}(\omega) {\bf G}^a(\omega) {\bf \Sigma}^{<}_L(\omega)- {\bf G}^r(\omega) {\bf \Sigma}_{ph}^<(\omega) {\bf G}^a(\omega) {\bf \Sigma_L^{>}}(\omega) \Big].
\eea
Note that in the above equations the retarded and advanced components of the Green's functions ${\bf G}^{r,a}$ are renormalized by the electron-photon interaction. 
We expand these functions following the Dyson equation, 
${\bf G}^{r,a}(\omega)= {\bf G}_0^{r,a}(\omega)+ {\bf G}_0^{r,a}(\omega) {\bf \Sigma}_{ph}^{r,a}(\omega) {\bf G}^{r,a}(\omega)$, and
organize the lowest order expression for the charge current in terms of the non-interacting ${\bf G}_0$. 
The elastic components of the current becomes $I_{el}=I_{el}^{(0)} + I_{el}^{(2)}$, with  
\be
I_{el}^{(0)} = e \int_{-\infty}^{\infty} \frac{d\omega}{2 \pi} {\rm Tr} \Big[ {\bf G}_0^r(\omega) {\bf \Gamma}_L(\omega) {\bf G}_0^a(\omega) {\bf \Gamma}_R(\omega) \Big] \big[f_L(\omega)-f_R(\omega)\big],
\label{Landauer-form}
\ee
and
\be
I_{el}^{(2)} = e \int_{-\infty}^{\infty} \frac{d\omega}{2 \pi} \Big({\rm Tr} \big[{\bf G}_0^r(\omega) {\bf \Sigma}^{r}_{ph}(\omega) {\bf \Gamma}_L(\omega) {\bf G}_0^a(\omega) {\bf \Gamma}_R(\omega) \big] + h.c.\Big) 
 \big[f_L(\omega)-f_R(\omega)\big].
\ee
%
The inelastic component is given by
\be
I_{inel}^{(2)} = e \int_{-\infty}^{\infty} \frac{d\omega}{2 \pi} {\rm Tr} \Big[{\bf G}_0^r(\omega) {\bf \Sigma}_{ph}^{>}(\omega) {\bf G}_0^a(\omega) {\bf \Sigma}_L^{<}(\omega)- {\bf G}_0^r(\omega) {\bf \Sigma}_{ph}^<(\omega) {\bf G}_0^a(\omega) {\bf \Sigma}_L^{>}(\omega) \Big].
\ee
In Fig.~\ref{current_fig} we display the coherent contribution to the current [Eq.~(\ref{Landauer-form})] and the total charge current, as well as  the photon induced (nonlinear) contribution, see panels (b) and (d). We examine the current 
as a function of $\epsilon$ for different values of inter-dot tunnelling $t$ and find that the coherent tunnelling component (solid line) dominates the total current in all cases. 
This coherent current is symmetric in $\epsilon$, with the maximum magnitude showing up at $\epsilon=0$, when electrons 
resonantly transfer through the energy-degenerate DQD setup. 
Turning on light-matter coupling induces an asymmetry in the current, with the nonlinear contribution 
showing clear signatures of photon-assisted transport, for both positive and negative detuning. 
The enhancement of the current correlates with the photon gain condition, as it appears around similar values of detuning. 
This further confirms that photons are generated in the cavity as a result of electron transfer processes between dots.
Note that the photon-induced contribution to the charge current can be extracted experimentally 
by measuring the total current under light-matter coupling, and then the current in the absence of coupling to the resonator.

\section{Conclusion}
\label{Summary}
To summarize, by using the Keldysh diagrammatic NEGF approach
we had investigated  the photonic (microwave) and the electronic properties of a non-equilibrium double quantum 
dot setup coupled to a microwave resonator.
The NEGF technique is well suited to describe this hybrid quantum system, given the parameter regime in which relevant experiments 
are performed: 
The method is valid for any value of the voltage bias, temperature, and electron-lead coupling.
While it accounts for the interaction between electrons and the electromagnetic field only in a perturbative manner, $g<\omega_0$, 
this restriction is appropriate for systems examined in experiments (see Table).
%

By looking at different observables, we had demonstrated that the DQD electronic system can serve as
a gain medium for the cavity photons when a {\it finite} source-drain bias was applied across the dots.
This effect, key for realizing photon-source quantum devices, was elucidated
by studying (analytically and numerically) the behavior of the
mean photon number, power spectrum and the spectral function of microwave photons, as well as the
transmission function and the phase response.
%
Specifically, we had interrogated the behavior of the 
photon number by sweeping the bias voltage, dot-lead hybridization and
detuning (bare energy difference between quantum dots), demonstrating
a significant enhancement of $\langle \hat n_c\rangle$ upon its coupling to the voltage-biased DQD system. 

The response of the cavity to signals has been investigated through its 
transmission coefficient and phase. We showed that the transmission could be greatly enhanced by tuning the dots' site-energies and the 
inter-dot tunneling energy, to 
satisfy a resonance condition with the frequency of the cavity.
A highly nontrivial observation has been that 
the transmission coefficient reaches a {\it maximum} at a certain {\it finite bias}
(Fig. \ref{trans_phase_bias}),
precisely once the imaginary part of the (electronic) charge susceptibility cancels out damping effects due to the (photonic) transmission lines,
$F^{''}_{el}=\kappa$.
Thus, the electronic system can be manipulated to minimize dissipation effects in the cavity, to improve its coherent properties. Our method and conclusions are therefore a major step forward towards Quantum Hamiltonian and Bath Engineering of hydrid quantum systems in general and QD-cQED systems in particular. 
%
These findings are principal for the realization of microwave amplifiers and masers, and more generally, quantum devices based on electron-induced 
photon gain and emission.


Besides properties of the cavity, we examined the behavior of the charge current across the DQD system as a function of level detuning.
Using the Meir-Wingreen formula to the lowest order in $g$, $O(g^2)$,
we identified photon-induced contributions to the current at both positive and negative detuning.

We emphasize that, for simplicity, in our treatment here we had assumed non-interacting electrons, by 
ignoring Coulombic repulsion energy between the dots' electrons.
Once included, this interaction would further affect gain/loss in the cavity, yet general features are expected to survive.
For example, the transmission coefficient should maintain its form [Eq. (\ref{eq:tdef})] when electron-electron (e-e)
interactions are accounted for; the electronic charge susceptibility $F^r_{el}$ will
be modified, to shift features and adjust the amplitude of transmission. 
Our approach---valid for any value of $\Gamma$--- should be contrasted with other common descriptions: 
In the Coulomb blockade regime (small  $\Gamma$ and an infinitely strong e-e repulsion)
the DQD can be conveniently described in the charge-state basis, eliminating the double occupancy (one electron on each dot)
state from the basis \cite{petta2014,Kulkarni2014}. It would be further interesting to incorporate, within our NEGF formalism, the presense of nanowire and bulk phonons~\cite{weber10}, influential in some quantum dot setups\cite{Kulkarni2014}. The existance of phonons can contain further interesting physics, for e.g, enhanced gain in signal \cite{pettamaser}, nonlinear functions such as diodes and transistors \cite{qdevice}. 

In future work we will apply the diagrammatic  approach and study more compound systems, 
for e.g., understanding the masing behavior that has been 
recently observed in a double double-quantum dot system \cite{pettamaserscience}.
On a more general front, we plan to extend the present formalism to spin systems coupled to light degrees of freedom \cite{kulkarniprx,kulberkeley}, 
further driven out of equilibrium by a source-drain voltage
realized in experiments on hybrid quantum systems \cite{fn1,fn2,pettamaserscience}.  
\section*{Acknowledgments}
We acknowledge discussions with J. Petta, T. Kontos, Y. Liu and M. Schiro. B.K.A and D.S were supported by the Natural Sciences and Engineering Research Council of Canada, the Canada Research Chair Program, and the Centre for Quantum Information and Quantum Control (CQIQC) at the University of Toronto. M.K. thanks the hospitality of the Chemical Physics Theory Group at the Department of Chemistry of the University of Toronto and the Initiative for the Theoretical Sciences (ITS) $-$ City University of New York Graduate Center where several interesting discussions took place during this work. He also gratefully acknowledges support from the Professional Staff Congress City University of New York award No.68193-0046. 
S. M. gratefully acknowledges the support of the National Science Foundation (NSF) through Grant No. CHE-1361516 and the U.S. Department of Energy (DOE).


\renewcommand{\theequation}{A\arabic{equation}}
\setcounter{equation}{0}  
\section*{Appendix A: Electron Green's functions}
In this Appendix we give expressions for different components of the non-interacting electronic Green's functions (dressed by the lead interaction) in the frequency domain. We write
\bea
{\bf G}_0^{r,a}(\omega)&=& \big[\omega {\bf I} - {\bf H}_S - {\bf \Sigma}^{r,a}(\omega)\big]^{-1},\nonumber \\
{\bf G}_0^{</>}(\omega)&=& {\bf G}_0^{r}(\omega) {\bf \Sigma}^{</>}(\omega) {\bf G}_0^{a}(\omega),
\eea
where ${\bf H}_S$ the Hamiltonian of the central DQD system  
\bea
&&{\bf H}_S=\begin{bmatrix}
                         \epsilon_1\,\, & \,\, t   \\
                          t\,\, & \,\,\epsilon_2 
\end{bmatrix} \eea
and ${\bf \Sigma}^{r,a,</>}(\omega)$ are different components of the total self-energy which is additive in the metallic leads $L$ and $R$, i.e., 
$
{\bf \Sigma}^{r,a,</>}(\omega)= {\bf \Sigma}_L^{r,a,</>}(\omega)+{\bf \Sigma}_R^{r,a,</>}(\omega).
$
Here ${\bf \Sigma}_L^{r,a}(\omega)= {\rm diag}(\mp\frac{i \Gamma_L(\omega)}{2},0)$, ${\bf \Sigma}_L^{<}(\omega)= {\rm diag} \big(i f_L(\omega) \Gamma_L , 0\big)$, ${\bf \Sigma}_L^{>}(\omega)= {\rm diag} \big(-i [1\!-\!f_L(\omega)] \Gamma_L , 0\big)$. In writing the components  $\Sigma_L^{r,a}(\omega)$ we ignore the real part, responsible for the renormalization of the DQD's energies. Similar expressions hold for the right lead self-energy, with $\Gamma_L \to \Gamma_R$ and $f_L (\omega)\to f_R(\omega)$.

\begin{figure}
\includegraphics[width=15cm]{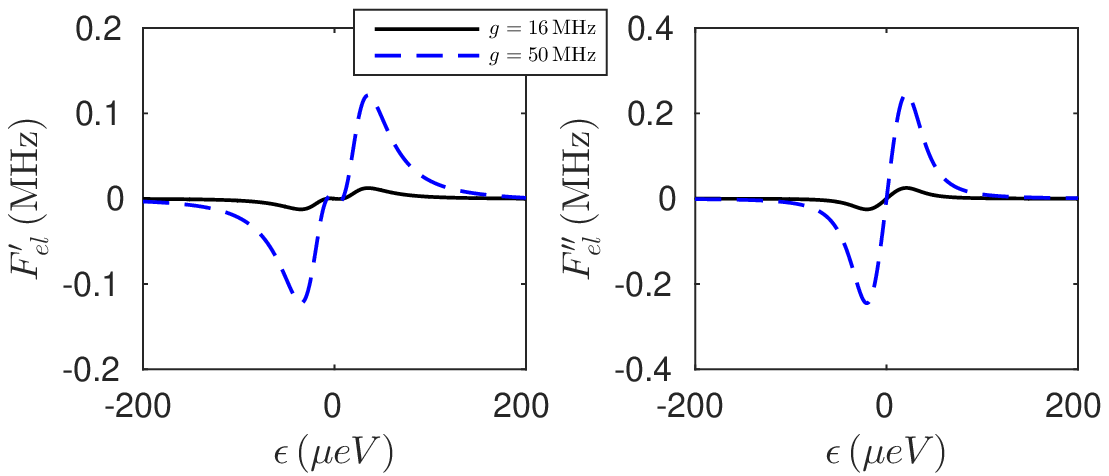}\\
\includegraphics[width=15cm]{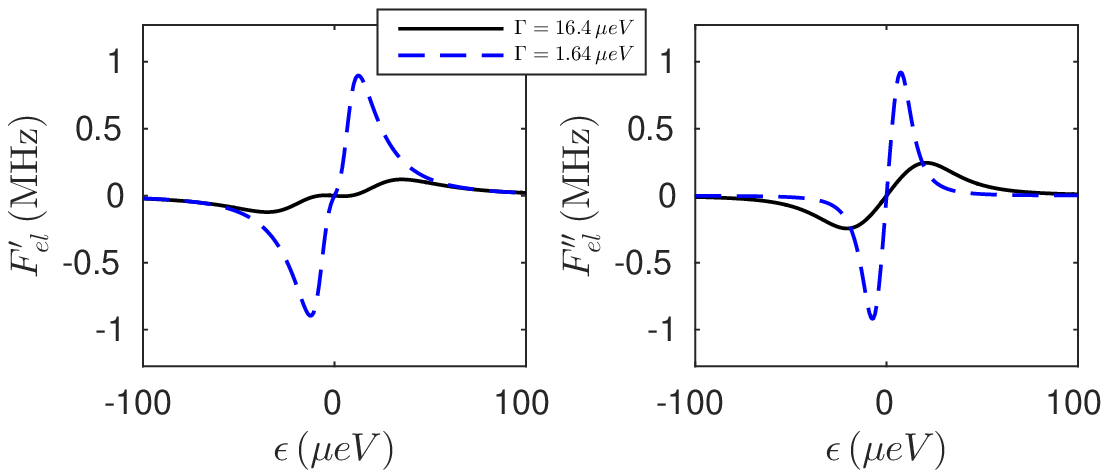}
\caption{(Color online) Real ($F_{el}^{'}$) and imaginary ($F_{el}^{''}$) components of the photon self-energy at the cavity mode frequency $\omega_0$ as a function of detuning. Parameters are $t\!=\!16.4\, \mu eV$, $\omega_0= 32.5\, \mu eV$, cavity loss rate $\kappa=0.01\, \mu eV \,(2.42\, {\rm MHz}) $, $\Delta \mu= 250 \,\mu eV$. (a) $\Gamma=16.56 \, \mu eV$, (b) $g \!=\! 50 \,{\rm  MHz}$.
Other parameters are the same as in table.~(\ref{tab:title}). } 
\label{self-energy}
\end{figure}
%
%
\renewcommand{\theequation}{B\arabic{equation}}
\setcounter{equation}{0} 
\section*{Appendix B: Expressions for bubble diagrams}
In this Appendix, we include expressions for the bubble diagrams. 
From Eq.~(\ref{eh-prop}) we receive the four different components
\bea
F_{el}^t (\omega) &=&- i \int_{-\infty}^{\infty} \frac{d\omega'}{2\pi} {\rm Tr} \Big[{\bf g} \, {\bf G}_{0}^{t}(\omega_+)\,{\bf g} \, {\bf G}_{0}^{t}(\omega_-) \Big], \nonumber \\
F_{el}^{\bar t} (\omega) &=&- i \int_{-\infty}^{\infty} \frac{d\omega'}{2\pi} {\rm Tr} \Big[{\bf g} \,{\bf G}_{0}^{{\bar t}}(\omega_+)\,{\bf g} \, {\bf G}_{0}^{{\bar t}}(\omega_-) \Big], \nonumber \\
F_{el}^{<} (\omega) &=&- i \int_{-\infty}^{\infty} \frac{d\omega'}{2\pi} {\rm Tr} \Big[{\bf g} \,{\bf G}_{0}^{<}(\omega_+)\,{\bf g} \, {\bf G}_{0}^{>}(\omega_-) \Big], \nonumber \\
F_{el}^{>} (\omega) &=&- i \int_{-\infty}^{\infty} \frac{d\omega'}{2\pi} {\rm Tr} \Big[{\bf g} \,{\bf G}_{0}^{>}(\omega_+)\,{\bf g} \, {\bf G}_{0}^{<}(\omega_-) \Big], \nonumber \\
\label{bubble-GF}
\eea
where $\omega_{\pm}= \omega'\pm \frac{\omega}{2}$. Note that $F_{el}^< (-\omega)=F_{el}^> (\omega)$. The sum and difference of the various components can be simplified following the above expressions, and using the causality condition for retarded and advanced Green's functions, in other words, $\int_{-\infty}^{\infty} d\omega' {\bf G}_{0}^{r,a}(\omega_{+})\, {\bf G}_{0}^{r,a}(\omega_{-})=0$ which gives 
\bea
F_{el}^t(\omega) + F_{el}^{\bar t}(\omega) &=& F_{el}^<(\omega) + F_{el}^{>}(\omega), 
\nonumber \\
F_{el}^t(\omega)-F_{el}^{\bar t}(\omega) &=& 2\, F_{el}^{'}(\omega), 
\nonumber\\
F_{el}^>(\omega)-F_{el}^{<}(\omega) &=& 2 \, i \,F_{el}^{''}(\omega),
\eea
where
\bea
F_{el}^{'}(\omega) &=& -i \,\int_{-\infty}^{\infty}
\frac{d\omega'}{8\pi} \Big\{ {\rm Tr} \Big[{\bf g} \, {\bf
G}_0^{k}(\omega_+)\, {\bf g}\, \big({\bf G}_{0}^{r}(\omega_-)\!+\! {\bf
G}_{0}^{a}(\omega_-)\big)\Big]\!+\! {\rm Tr}\Big[ {\bf g}\,{\bf
G}_{0}^{k}(\omega_-)\, {\bf g}\, \big({\bf G}_{0}^{r}(\omega_+)\!+\! {\bf
G}_{0}^{a}(\omega_+)\big) \Big]\Big\}, \nonumber \\
F_{el}^{''}(\omega) &=& \int_{-\infty}^{\infty} \frac{d\omega'}{8\pi}
\Big\{ {\rm Tr} \Big[{\bf g} \, {\bf G}_0^{k}(\omega_+)\, {\bf g}\,
\big({\bf G}_{0}^{r}(\omega_-)\!-\!{\bf G}_{0}^{a}(\omega_-)\big)\Big]-{\rm
Tr}\Big[ {\bf g}\,{\bf G}_{0}^{k}(\omega_-)\, {\bf g}\, \big({\bf
G}_{0}^{r}(\omega_+)\!-\!{\bf G}_{0}^{a}(\omega_+)\big)\Big]\Big\}.
\nonumber \\
\label{real-im}
\eea 
Recall that $F^{'}_{el}(\omega)= Re[F^r(\omega)]$, $F^{''}_{el}(\omega)= Im[F^r(\omega)]$, $g={\rm diag}(g_1, g_2)$ and ${\bf G}_{0}^{k}={\bf G}_{0}^{<}+ {\bf G}_{0}^{>}$, the Keldysh component. The nontrivial bias dependence in the real and imaginary parts of $F_{el}^{r}$ enters through the ${\bf G}_{0}^{k}$ component. The $F_{el}^{'}(F_{el}^{''})$ terms depend on the reactive (absorptive) part of the electronic Green's function. Note that, at equilibrium, detailed balance condition for the Green's functions, given as ${\bf G}_{0}^{>}(\omega)=-e^{\beta(\omega-\mu)} {\bf G}_{0}^{<}(\omega)$,  immediately implies $F_{el}^{>}(\omega)= e^{\beta \omega} F_{el}^{<}(\omega)$.
In Fig.~\ref{self-energy} we display $F_{el}^{'}(\omega_0)$ and $F_{el}^{''}(\omega_0)$ as a function of detuning for various $g$ and $\Gamma$. Increasing $g$ and reducing $\Gamma$ leads to a significant electron-photon interaction, thereby leading to large values for the elements of $F_{el}(\omega_0)$.
\vspace{-0.5cm}
\section*{Korringa-Shiba relation}
\vspace{-0.5cm}
The gain in the transmission is determined by the real and imagniary components of $F_{el}(\omega)$. Therefore understanding the relation between these components is important. Here we give compact expressions for $F_{el}^{',''}(\omega)$ at the lowest order of $\omega$ for zero bias ($\mu_L\!=\!\mu_R\!=\!\mu$). In the simpler setup of the Anderson impurity model, it has been shown that in this limit the $F_{el}^{'}(\omega)$ and $F_{el}^{''}(\omega)$ components are not independent but rather related via the Korringa-Shiba relation \cite{shiba_korringa, LeHur_shiba,simon16}. An experimental test of this relation based on a microwave cavity setup has also been proposed in Ref.~[\onlinecite{cottetQED}]. Below we discuss this realation for our DQD setup. From Eq.~(\ref{real-im}) in the zero temperature limit we obtain  
\bea
F^{'}_{el}(\omega\!\to \!0, \Delta \mu\!=\!0)&=&{i} \,\int_{-\infty}^{\mu} \frac{d\omega'}{2\pi} {\rm Tr} \big[{\bf g}  \, {\bf G}_0^r(\omega') \, {\bf g} \, {\bf G}_0^r(\omega')-{\bf g}  \, {\bf G}_0^a(\omega') \, {\bf g} \, {\bf G}_0^a(\omega') \big], \nonumber \\
F^{''}_{el}(\omega \!\to\! 0, \Delta \mu\!=\!0)&=&-\omega \int_{-\infty}^{\mu} \frac{d\omega'}{2\pi} {\rm Tr} \Big[{\bf g}  \, {\bf A}_{el}(\omega') \, {\bf g} \,  \frac{\partial {\bf A}_{el}(\omega')}{\partial \omega'} \Big] = -\frac{\omega}{4 \pi} {\rm Tr} \Big[{\bf g} \, {\bf A}_{el}(\mu) \, {\bf g} \, {\bf A}_{el}(\mu) \Big].\nonumber \\
\eea
where ${\bf A}_{el}(\omega)= i \big[{\bf G}_0^r(\omega)-{\bf G}_0^r(\omega)\big]$ is the spectral function for the DQD system.
In the wide band limit ($\Gamma(\omega)=\Gamma$), using the relation  
\be
\frac{\partial {\bf G}^{r,a}_0(\omega)}{\partial \omega} \!=\! - \,{\bf G}^{r,a}_0(\omega)\, {\bf G}^{r,a}_0(\omega)
\ee
one can further simplify the real and imaginary components of $F_{el}(\omega)$. As an example, for $g_1\!=\!g_2\!=\!g$, ${\bf g}$ is proportional to an identity matrix. The above expressions then reduce to 
\bea
F^{'}_{el}(\omega \!\to\!0, \Delta \mu\!=\!0)&=& -\frac{g^2}{2\pi} {\rm Tr}\big[{\bf A}_{el}(\mu)\big], \nonumber \\
F^{''}_{el}(\omega \!\to\!0, \Delta \mu\!=\!0)&=& -\frac{g^2\, \omega}{4\pi} {\rm Tr}\big[{\bf A}_{el}^2(\mu)\big].
\eea
Note that the differential conductance $dI/d(\Delta \mu)$ is often noted to be proportional to the spectral function. This indicates that $F^{'}_{el}$ and $dI/d(\Delta \mu)$ may show similar features, as confirmed in a recent experiment \cite{kontoskondo}.

In contrast, for a {\it single} site coupled to the photon mode i.e., $g_1=g, g_2=0$ or vice versa, we recover the standard Korringa-Shiba relation
\be
F^{''}_{el}(\omega\!\to\!0,\Delta \mu\!=\!0)=-\frac{\omega \,\pi}{g^2} \,\big[F^{'}_{el}(\omega \!\to\! 0, \Delta \mu\!=\!0)\big]^2.
\ee
\vspace{-1cm}
\bibliographystyle{apsrev}
\bibliography{references}

\end{document}